\documentclass[%
 reprint,
nofootinbib,
 amsmath,amssymb,
 aps,
 superscriptaddress,
prd,
]{revtex4-2}

\usepackage{graphicx}
\usepackage{dcolumn}
\usepackage{xcolor}
\usepackage{bm}
\usepackage{multirow}
\usepackage{hyperref}
\usepackage{xspace}

\usepackage[normalem]{ulem}
\newcommand{\um}{\texttt{UniverseMachine}}
\newcommand{\sides}{\texttt{SIDES} }

\newcommand{\lir}{L_{\rm IR}}
\newcommand{\skyline}{\texttt{SkyLine}\xspace}
\newcommand{\agora}{\texttt{Agora}\xspace}

\newcommand{\btheta}{\boldsymbol{\theta}}
\newcommand{\universemachine}{\texttt{UniverseMachine}\xspace}

\makeatletter
\newcommand*{\citelinktext}[2]{%
  \hyper@@link[cite]{}{cite.#1}{#2}}
\makeatother
\newcommand{\Maniyar}{\citelinktext{Maniyar_2023}{M23}\xspace}

\begin{document}

\preprint{APS/123-QED}

\title{On the contributions of extragalactic CO emission lines to ground-based CMB observations}
\author{Nickolas Kokron}
\email{kokron@astro.princeton.edu}
\affiliation{Department of Astrophysical Sciences, Princeton University, 4 Ivy Lane, Princeton, NJ, 08544, USA}
\affiliation{School of Natural Sciences, Institute for Advanced Study, 1 Einstein Drive, Princeton, NJ, 08540, USA}
\author{José Luis Bernal}
\affiliation{Instituto de Física de Cantabria (IFCA), CSIC-Univ. de Cantabria, Avda. de los Castros s/n, E-39005 Santander, Spain}
\author{Jo Dunkley}
\affiliation{Department of Astrophysical Sciences, Princeton University, 4 Ivy Lane, Princeton, NJ, 08544, USA}
\affiliation{Joseph Henry Laboratories of Physics, Jadwin Hall, Princeton University, Princeton, NJ, 08540, USA}
\date{\today}

\begin{abstract}
We investigate the potential of CO rotational lines at redshifts $z\sim 0-6$ being an appreciable source of extragalactic foreground anisotropies in the cosmic microwave background. Motivated by previous investigations, we specifically focus on the frequency bands and small scales probed by ground-based surveys. Using an empirical parameterization for the relation between the infrared luminosity of galaxies and their CO line luminosity, conditioned on sub-mm observations of CO luminosity functions from $J=1$ to $J=7$ at $\nu = \{100,250\}$ GHz, we explore how uncertainty in the CO luminosity function translates into uncertainty in the signature of CO emission in the CMB. {We find that at $\ell = 3000$ the amplitude of the CO cross-correlation with the CIB could be detectable in an ACT-like experiment with 90, 150 and 220 GHz bands, even in the scenarios with the lowest amplitude consistent with sub-mm data}. We also investigate, for the first time, the amplitude of the CO$\times$CIB correlation between different frequency bands and find that our model predicts that this signal {could be the second-largest extragalactic foreground at certain wavelengths, behind the CIB cross-frequency spectrum}. This implies current observations can potentially be used to constrain the bright end of CO luminosity functions, which are difficult to probe with current sub-mm telescopes due to the small volumes they survey. Our findings corroborate past results and have significant implications in template-based searches for CMB secondaries,  such as the kinetic Sunyaev Zel'dovich effect, using the frequency-dependent high-$\ell$ TT power spectrum. 
\end{abstract}

\maketitle


\section{Introduction}
In recent years, high resolution ground-based cosmic microwave background (CMB) surveys, such as the Atacama Cosmology Telescope (ACT~\cite{2020JCAP...12..047A}) and the South Pole Telescope (SPT~\cite{10.1063/1.3292381, Austermann_2012}), have significantly extended the range of angular multipoles over which the temperature and polarization anisotropies of the CMB have been measured to high precision. Upcoming surveys such as the Simons Observatory~\cite{Ade_2019}, as well as the Prime-Cam on the Fred Young Submillimeter Telescope~\cite{CCAT_Prime_Collaboration_2022} will further extend this range. As the primary CMB fluctuations are exponentially damped at large $\ell$, the dominant sources of power at small scales are additional components which source photons at CMB frequencies -- secondary anisotropies of the CMB and extragalactic forerounds. These components are typically assumed to be uncorrelated with the primordial CMB temperature, and are generally written as linear contributions to the CMB temperature decrement at a sky direction $\btheta$
\begin{align}
    \Delta T^{\rm CMB} (\btheta) &= \Delta T^{\rm Prim}(\btheta) + \Delta T^{\rm tSZ}(\btheta) + \Delta T^{\rm kSZ}(\btheta) \nonumber \\
    &+ \Delta T^{\rm CIB} (\btheta) +\Delta T^{\rm RPS}\cdots, 
\end{align}
where the superscripts tSZ and kSZ 
refer to CMB secondary anisotropies like the thermal and kinetic Sunyaev-Zel'dovich effects, and CIB and RPS denote the cosmic infrared background and contamination from radio point sources which dominate at low frequencies, respectively. The power spectra of each of these components, and their cross correlations, have differing angular dependence and scaling with frequency -- templates of these signals have their amplitudes constrained through joint analyses of CMB maps observed at different frequencies~\cite{Reichardt_2012,Dunkley_2013, PlanckXIII, Reichardt_2021}. \par
\begin{figure}
    \centering
    \includegraphics[width=\columnwidth]{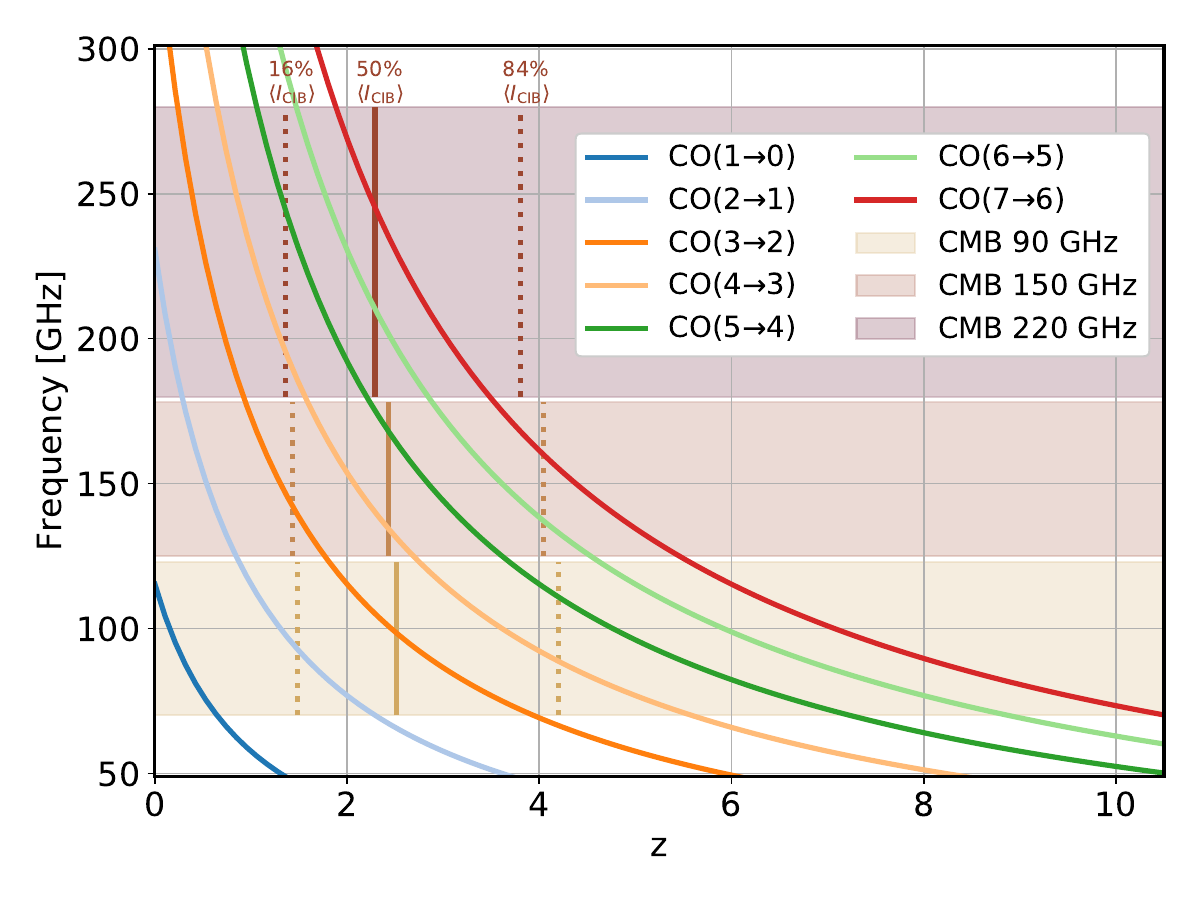}
    \caption{The redshift-observed frequency relation for the CO rotational ladder, compared to frequency bands representative of ground-based CMB surveys. The vertical lines bracket the redshift range from which the average CIB intensity in that band receives the 16\%, 50\% and 84\% of its contribution. Note the significant overlap between several CO lines and the peak of CIB emission.}
    \label{fig:freqs}
\end{figure}

Beyond the additional components of CMB photons discussed above, 
emission from rotational transitions of CO molecules contained in star-forming galaxies will also source photons at CMB frequencies. Indeed, a photon sourced from the CO$(J\to J-1)$ transition at redshift $z_{\rm em}$ will reach a CMB detector at 
\begin{equation}
\label{eqn:nuobsofz}
    \nu_{\rm obs} = \frac{J \times 115.271}{1 + z_{\rm em}} \,{\rm GHz}\,.
\end{equation}
Thus, for frequency bands of ground-based experiments, 
there is a wide swath of redshifts over which excited transitions fall into observed frequencies, especially at $\nu_{\rm obs} \sim 90, 150, 220$ GHz. In Fig.~\ref{fig:freqs} we show a visual illustration of Eqn.~\ref{eqn:nuobsofz} -- how the transitions from $J=1$ to $J=7$ redshift into three of the frequency bands of a typical ground-based CMB experiment. Specifically, the shaded bands show frequencies with significant transmission from passbands of an ACT-like experiment. When generating maps we use actual passbands from AdvACT detectors. We also show the redshift range over which 16-84\% of the intensity of the CIB is emitted, according to the CIB model of the \agora simulation (which we will introduce shortly). It is clear that for higher frequency bands there are many lines sourced at $z\sim 1-3$, the peak epoch of cosmic star-formation-rate density~\cite{Universemachine}, that are redshifted into the frequencies measured by CMB experiments. \par 
The 
potentially sizable contributions from redshifted CO transitions into CMB bands were originally noted by Ref.~\cite{Righi_2008}. However, the authors correctly noticed that as the spectral energy distribution of a molecular transition is closely approximated by a delta function, this signal was heavily suppressed by the width of the bandpass (by a factor of $\sim 1/\Delta \nu$). The auto-correlation would be further suppressed by $(\Delta \nu)^{-2}$ and it was suggested this signal would be challenging to detect unless radio experiments with very fine frequency resolution were built -- now known as line-intensity mapping surveys~\cite{Bernal:2022jap}. However, it was recently pointed out in Ref.~\cite{Maniyar_2023} (hereinafter M23) that, even though the auto-correlation of this signal is heavily suppressed for CMB surveys, its cross-correlation with the CIB could be significant. The CO$\times$CIB signal in their analysis was potentially comparable to the amplitude of the late-time kSZ effect, which had recently been claimed to be detected at 
$\sim 3\sigma$-significance in Ref.~\cite{Reichardt_2021}.\par 
Motivated by the fact that a potentially bright extragalactic foreground could be detectable in current ground-based CMB experiments and at the same time may have been biasing template-based and component separation analyses, in this work we produce independent predictions for the frequency-dependent power spectrum of CO fluctuations as well as their cross-correlations with the CIB. This is achieved by leveraging the \agora~\cite{omori2022agora}+\skyline\cite{skyline} cosmological N-body simulation suite of the multi-component sky. The fiducial \agora simulation has produced highly accurate CIB maps consistent with Planck, ACT and SPT multi-frequency data which are derived from an empirical star-formation model~\cite{Universemachine} coupled to a prescription to map star formation to IR luminosity. We use the same large-scale-structure scaffolding and star formation to derive CO luminosities. This allows us to produce full-sky maps of line emission across many transitions of the CO rotational ladder, and study their multi-frequency impact on CMB maps. The use of the full-sky \agora simulation allows us to probe this emission at a larger volume than was able to be done in \Maniyar. \par 
Additionally, in this work we go beyond standard assumptions for the CO emission ladder and calibrate the individual transitions to observed data at 100 and 250 GHz, near frequencies where ground-based CMB observatories operate. Current measurements of CO luminosity functions, due to their small fields of view, do not sample the entirety of the luminosity function with high precision. We therefore will use the flexibility of our empirical model to attempt to quantify the uncertainty associated with the impact of CO transitions on CMB maps -- thus answering the question \emph{how much impact on CMB observables is allowed, conditioned on CO luminosity functions consistent with data?}.\par 
This paper is structured as follows: in \S~\ref{sec:sims} we discuss the techniques adopted to simulate IR and CO emission from the underlying large-scale structure of a cosmological halo lightcone, and how we use this catalog to  construct maps of CO emission for CO transitions from $(1\to0)$ to $(7\to6)$ for three bandpasses representing an ACT-like CMB survey. In \S~\ref{sec:cells} we present measurements of angular power spectra of the CO maps produced in this work, as well as cross-correlations with CIB maps. We compare these power spectra to the CIB auto-spectrum as well as the imprint of the late-time kSZ effect. We also discuss power spectra across different frequency bins for combinations of CO$\times$CIB. In \S~\ref{sec:discussion} we present some qualitative discussion on the amplitude of the observed correlations in the previous section. We also compare with predictions for similar observables reported in \Maniyar. We conclude in \S~\ref{sec:conclusions} and discuss future directions that we believe are particularly pertinent in light of the results presented in this publication. 
\section{Empirical simulations of CMB secondaries and extragalactic foregrounds}
\label{sec:sims}
Quantifying the amplitude of extragalactic foregrounds and secondary anisotropies is a challenging task. From a halo model modeling perspective, it requires an in-depth understanding of the distribution of halos and their electron pressure profiles (for tSZ), small-scale density profiles (for kSZ), or a detailed understanding of the connection between halo masses and radio / IR luminosity; all within a cosmological context across a wide range of redshift. While halo model approaches have been successful at analytically modeling many of these secondaries (see, e.g., Ref.~\cite{bolliet2023classsz}), building halo models for CO emission is a subject in its infancy, and mainly dedicated to modeling 
line-intensity mapping observations (see e.g.,~\cite{Bernal_2019, 2019ApJ...887..142S}). Effectively, a halo model for CO emission would have to capture the evolution of the connection between halo mass and the luminosity of a specific CO $J$ transition between $0 < z \lesssim 5$, as well as the correlations between different transitions within a specific halo~\cite{Schaan_2021}.\par 
An alternative approach is to rely on empirical observations of CO luminosity and how it correlates with other astrophysical observables. These empirical data may be used to assign CO luminosities to halos in large N-body simulations, which try to capture the underlying correlations between halo mass and the astrophysical observable which scales with CO luminosity. In the context of CMB secondaries and extragalactic foregrounds these empirically-calibrated simulations have been generated using peak-patch algorithms~\cite{Stein_2020} or by adopting more realistic models which attempt to model star-formation histories for dark matter halos in a fully nonlinear cosmological $N$-body simulation, such as in the \sides or \agora multi-component sky simulations~\cite{B_thermin_2017, B_thermin_2022, omori2022agora}. \par 
In this work, we have elected to use the simulation suite underpinning \agora observables to empirically assign CO luminosities to underlying halos. The shared distribution of halo masses, stellar masses, and star formation rates (SFR) will ensure our CO maps correlate with other extragalactic foregrounds and CMB secondaries in \agora, given our model. The large-scale structure underpinning \agora is the dark matter distribution from the \texttt{MultiDark Planck 2} cosmological simulation~\cite{MDPL2}, which has a volume of V=1 
 $[{\rm Gpc}\,h^{-1}]^3$ and $N=(3840)^3$ particles, resulting in a mass resolution of $M_p = 1.51 \times 10^{9} M_\odot h^{-1}$. The \texttt{Rockstar} phase-space halo finder~\cite{Rockstar} is applied to the resulting particle catalogs, and the empirical \texttt{UniverseMachine} model~\cite{Universemachine} is then used to assign SFR, stellar masses and star-formation histories to every halo and sub-halo in the catalog. \agora then builds full-sky lightcones from these halo catalogs at different redshifts.
 \par
 The original \agora publication showed that using stellar masses and SFR derived from \um, the resulting lightcones reproduce the correlations of fields which depend on these quantities, such as anisotropies of the CIB, in a way consistent with current data. The fiducial realization of \agora is also tuned to describe summary statistics related to the tSZ effect, the kSZ effect, radio-bright galaxies, cosmic shear, CMB lensing, and more.  We refer to Ref.~\cite{omori2022agora} for a more detailed description of how each individual component of \agora is constructed. We note, however, that in this study we use maps of \emph{unlensed} CIB anisotropies as we anticipate the impact of CMB lensing to be subleading in our observables, and accurately lensing the CO emission we wish to simulate is a computationally challenging procedure~\cite{Schaan_2018}. \par

\begin{table}[t!]
\centering
\resizebox{0.7\columnwidth}{!}{
\begin{tabular}{|l|ccc|}
\hline
\multirow{3}{*}{\begin{tabular}[c]{@{}l@{}}CO\\ 
transitions\end{tabular}}  & \multicolumn{3}{c|}{Frequency band} \\ \cline{2-4} 
 & \multicolumn{1}{c|}{90 GHz} & \multicolumn{1}{c|}{150 GHz} & 220 GHz \\ \cline{2-4} 
 & \multicolumn{3}{c|}{Redshift range} \\ \hline
CO(1-0) & \multicolumn{1}{c|}{0-0.5} & \multicolumn{1}{c|}{-} & - \\ \hline
CO(2-1) & \multicolumn{1}{c|}{1.1-2.0} & \multicolumn{1}{c|}{0.3-0.9} & 0-0.2 \\ \hline
CO(3-2) & \multicolumn{1}{c|}{2.1-3.5} & \multicolumn{1}{c|}{1.0-1.8} & 0.2-0.8 \\ \hline
CO(4-3) & \multicolumn{1}{c|}{3.1-5.0} & \multicolumn{1}{c|}{1.7-2.7} & 0.7-1.5 \\ \hline
CO(5-4) & \multicolumn{1}{c|}{4.1-6.5} & \multicolumn{1}{c|}{2.4-3.6} & 1.1-2.1 \\ \hline
CO(6-5) & \multicolumn{1}{c|}{5.2-8.0} & \multicolumn{1}{c|}{3.0-4.6} & 1.5-2.7 \\ \hline
CO(7-6) & \multicolumn{1}{c|}{6.2-9.5} & \multicolumn{1}{c|}{3.7-5.5} & 1.9-3.3 \\ \hline
\end{tabular}}
\caption{Redshift ranges for each CO transition sampled by frequency bands of ground-based observatories, assuming for ACT-like bandpasses. Note that all frequencies receive significant contributions from CO emission at the redshifts over which the contribution of star formation to the CIB peaks, which is between redshifts $1<z<3$.}
\label{tab:zs}
\end{table}

\subsection{Infrared luminosities of halos}
 \label{subsec:lir}
 Our scheme to sample empirical CO luminosities in dark matter halos will leverage the well-studied connection between CO luminosity and IR luminosity in CO-bright galaxies (see e.g.,~\cite{Carilli:2013qm, 2016ApJ...829...93K}), and thus we begin by assigning IR luminosities to the halos (including sub-halos) in our lightcone. It is also important to  ensure that our CO luminosities correlate with \agora CIB maps correctly, and thus we will assign IR luminosities to halos in the same way as done in \agora. We briefly recap their approach but refer to Ref.~\cite{omori2022agora} for full details. \par 
 A modified Kennicutt relation is used to sample an IR luminosity for every halo according to its SFR and stellar mass ($M_*$), following 
 \begin{equation}
 \label{eqn:kennicutt}
     L_{\rm IR} = \frac{\rm SFR}{K_{\rm IR} + K_{\rm UV} {\rm IRX}^{-1}(M_*)},
 \end{equation}
 where $K_{\rm UV, IR}$ are coefficients with units of $[K_{\rm UV, IR}] = M_\odot {\rm yr}^{-1} L_{\odot}^{-1}$ taken from~\cite{omori2022agora}\footnote{Who originally reference Ref.~\cite{2012ARA&A..50..531K} with a correction for initial mass functions taken from Ref.~\cite{MadauDickinson}.}, and ${\rm IRX}$ is the \emph{IR excess} quantifying the ratio between ultraviolet and IR luminosity for a given galaxy. It is parameterized as a power-law in stellar mass~\cite{Bouwens2020},
 \begin{equation}
     {\rm IRX} = \left ( \frac{M_*}{M_s} \right )^{\alpha_{\rm IRX}},
 \end{equation}
with $M_s$ and $\alpha_{\rm IRX}$ values taken from Ref.~\cite{omori2022agora}. This results in a close-to-linear relation between IR luminosity and SFR for high SFR, but showing a deficit for low SFR values and overall deviations at high redshifts (see Fig.5 in Ref.~\cite{omori2022agora}). 
\universemachine  assigns two types of SFR for a halo, `true' and `observed', and we use `observed' SFRs in Eqn.~\ref{eqn:kennicutt}. Since the `observed' SFR and $M_*$ drawn by \universemachine already include scatter, we do not assign additional scatter and instead evaluate Eqn.~\ref{eqn:kennicutt} directly to assign $L_{\rm IR}$ to a halo. This is consistent with what was implemented in \agora. \par 
The assigned IR luminosity is interpreted as a bolometric value\footnote{Specifically for the SED in the frequency range of $\nu = [1,4 \times 10^4]$ GHz.}, and in order to compute CIB anisotropies a spectral energy distribution (SED) must be assigned to every halo. \agora adopts a modified black-body SED with a transition frequency into a power law, with slope determined by the dust temperature, $T_d$, on a halo-by-halo basis. This dust temperature may be estimated from the SFR and stellar mass of each halo, giving a self-consistent solution for IR luminosity and spectral index arising from only the halo mass, stellar mass and SFR of that halo. This final SED is redshifted according to the halo's redshift in the underlying lightcone, and the resulting k-corrected flux~\cite{hogg2000distance} is integrated against the bandpass of a CMB experiment in order to compute the CMB temperature decrement induced by IR photons coming from an individual galaxy for that experiment.\par 
This is the scheme by which \agora constructs maps of diffuse CIB emission that contaminates CMB maps. Following this scheme to derive other observable quantities ensures that their cross-covariance is taken into account in the simulation suite.
\begin{table}[t]
\centering
\resizebox{\columnwidth}{!}{%
\begin{tabular}{|r|ccc}
\hline
\multicolumn{1}{|l|}{} & \multicolumn{3}{c|}{$\alpha_{\rm IR}$; $\beta_{\rm IR}$; Reference} \\ \cline{2-4} 
\multicolumn{1}{|c|}{\begin{tabular}[c]{@{}c@{}}CO($J\rightarrow J-1$)\\ transitions\end{tabular}} & \multicolumn{3}{c|}{Frequency band} \\ \cline{2-4} 
\multicolumn{1}{|l|}{} &  \multicolumn{1}{c|}{90} & \multicolumn{1}{c|}{150} & \multicolumn{1}{c|}{220}\\ \hline

 &  \multicolumn{1}{c|}{} & \multicolumn{1}{c|}{-} & \multicolumn{1}{c|}{-} \\
1-0 &  \multicolumn{1}{c|}{1.27; 1; \cite{2016ApJ...829...93K}} & \multicolumn{1}{c|}{-} & \multicolumn{1}{c|}{-} \\
 &  \multicolumn{1}{c|}{} & \multicolumn{1}{c|}{-} & \multicolumn{1}{c|}{-} \\ \hline
 
 & \multicolumn{1}{c|}{0.91; 2.51} & \multicolumn{1}{c|}{0.98; 1.55;} & \multicolumn{1}{c|}{} \\
2-1 &  \multicolumn{1}{c|}{\begin{tabular}[c]{@{}c@{}}0.97; 2.04; \cite{2020ApJ...902..110D,  Boogaard:2023whs}\end{tabular}} & \multicolumn{1}{c|}{1.03; 1.30; (*)} & \multicolumn{1}{c|}{1.11; 0.6; \cite{2016ApJ...829...93K}} \\
 &  \multicolumn{1}{c|}{1.19; 0.03} & \multicolumn{1}{c|}{1.15; 0.40} & \multicolumn{1}{c|}{} \\ \hline
 
 &  \multicolumn{1}{c|}{0.99; 1.65} & \multicolumn{1}{c|}{1.06; 1.02} & \multicolumn{1}{c|}{1.18; -0.56} \\
3-2 &  \multicolumn{1}{c|}{1.06; 1.24; \cite{2020ApJ...902..110D, Boogaard:2023whs}} & \multicolumn{1}{c|}{1.06; 1.35; (*)} & \multicolumn{1}{c|}{1.06; 0.85; \cite{2020ApJ...902..110D}} \\
 & \multicolumn{1}{c|}{1.31; -0.91} & \multicolumn{1}{c|}{1.3; -0.65} & \multicolumn{1}{c|}{0.94; 2.31} \\ \hline
 
 & \multicolumn{1}{c|}{1.92; -8.04} & \multicolumn{1}{c|}{1.55; -3.72} & \multicolumn{1}{c|}{1.04; 1.31} \\
4-3 &  \multicolumn{1}{c|}{3.20; -20.67; \cite{2020ApJ...895...81R, Boogaard:2023whs}} & \multicolumn{1}{c|}{1.83; -6.20; (*)} & \multicolumn{1}{c|}{1.02; 1.69; \cite{2020ApJ...902..110D}} \\
 & \multicolumn{1}{c|}{3.60; -23.94} & \multicolumn{1}{c|}{1.98; -7.20} & \multicolumn{1}{c|}{1.02; 2.04} \\ \hline
 
 &  \multicolumn{1}{c|}{1.00; 1.08} & \multicolumn{1}{c|}{1.03; 1.58} & \multicolumn{1}{c|}{1.07; 1.29} \\
5-4 & \multicolumn{1}{c|}{1.10; 0.57; \cite{Boogaard:2023whs}} & \multicolumn{1}{c|}{1.07; 1.51; (*)} & \multicolumn{1}{c|}{1.05; 1.79; \cite{2020ApJ...902..110D}} \\
 & \multicolumn{1}{c|}{1.28; -1.22} & \multicolumn{1}{c|}{1.29; -0.55} & \multicolumn{1}{c|}{1.30; -0.55} \\ \hline
 
 &  \multicolumn{1}{c|}{ 1.22; -0.75} & \multicolumn{1}{c|}{ 0.94; 2.56} & \multicolumn{1}{c|}{1.06; 1.54} \\
6-5 &  \multicolumn{1}{c|}{ 1.42; -2.05; (**)} & \multicolumn{1}{c|}{ 1.02; 1.98; (**)} & \multicolumn{1}{c|}{1.06; 1.74; \cite{2020ApJ...902..110D}} \\
 & \multicolumn{1}{c|}{ 1.62; -3.32 } & \multicolumn{1}{c|}{ 1.11; 1.52} & \multicolumn{1}{c|}{1.16; 0.95} \\ \hline
 
 &  \multicolumn{1}{c|}{ 1.36; -1.65} & \multicolumn{1}{c|}{ 0.88; 3.23} & \multicolumn{1}{c|}{1.0; 2.2} \\
7-6 &  \multicolumn{1}{c|}{ 1.51; -2.25; (**)} & \multicolumn{1}{c|}{ 0.98; 2.37; (**)} & \multicolumn{1}{c|}{1.03; 2.06; \cite{2020ApJ...902..110D}} \\
 &  \multicolumn{1}{c|}{ 1.67; -3.32} & \multicolumn{1}{c|}{ 1.08; 1.76} & \multicolumn{1}{c|}{1.07; 1.90} \\ \hline
\end{tabular}%
}
\caption{Values for the $\alpha_{\rm IR}$ and $\beta_{\rm IR}$ parameters appearing in the $L_{\rm IR}-L'_{J}$ relation. We show the values used for each transition and frequency band (corresponding to different redshift ranges, according to Table~\ref{tab:zs}), fit to luminosity functions measured in overlapping redshift ranges (references indicated in the table). Values obtained from interpolating and extrapolating the redshift evolution of the $L'$ luminosity functions are denoted by (*) and (**), respectively. Every transition-frequency combination has three values shown, corresponding to (from top to bottom) the `shallow', `fiducial', and 'steep' luminosity functions discussed in the text.}
\label{tab:alphabeta}
\end{table}
\subsection{CO Luminosity functions}
\label{subsec:COLF}
The CO luminosities of every halo, combined with their IR luminosity, are the building blocks of the extragalactic CMB foregrounds we wish to study in this work. Given a realization of the correlated extragalactic sky, we derive CO luminosities across all relevant transitions, on a halo-by-halo basis, using \skyline\footnote{\href{https://github.com/kokron/skyline}{https://github.com/kokron/skyline}}~\cite{skyline}, which implements an updated version of the CO painting scheme originally proposed in Ref.~\cite{Li_2016}.  We assume that CO luminosities can be parameterized using power-law relations, starting from IR luminosities that are computed using the scheme described in the previous section. Thus, we relate them to the observed CO luminosity for a given transition $L_{\rm CO}^{J \to J-1}$ with\footnote{These logarithms are in base 10, and this convention will hold for the rest of the paper.} 
\begin{equation}
\label{eqn:loglirco}
    \log \lir = \alpha_{J,\rm IR} \log {L^\prime}_{\rm CO}^{J \to J-1} + \beta_{J,\rm IR}, 
\end{equation}
where $\alpha_{J, \rm IR}$ and $\beta_{J, \rm IR}$ are the power-law slope and amplitude for that transition. This relation is not deterministic --values of $L'_{\rm CO}$ are sampled from this mean relation and we assume a fixed mean-preserving lognormal scatter of $\sigma_{\rm CO} = 0.2$ for all transitions and redshifts. The $L^\prime_{\rm CO}$ luminosity, in units of pseudo-luminosity, is related to physical units through
\begin{equation}
\label{eqn:pseudoLconvert}
    L_{\rm CO} = 4.9\times 10^{-5}L'_{\rm CO}J^3 \frac{L_\odot}{{\rm K\,km\, s^{-1}\,pc^2}}\,.
\end{equation}
A commonly adopted assumption in the literature is that the ratio between line luminosities at different transitions is constant for all redshifts (see e.g.,~\cite{Sun_2021}). CO emission is then modeled assuming $\alpha_{J,\rm IR} = \alpha_{0, \rm IR}$, thus writing that 
\begin{equation}
\label{eqn:lcolir}
     \log {L^\prime}_{\rm CO}^{J \to J-1} = \frac{\log \lir - \beta_{\rm IR}}{\alpha_{\rm IR}} + r_J,
\end{equation}
where $r_J$ sets the relative line luminosities for each transition across all redshifts. The values of $\alpha, \beta$ are commonly calibrated to low-redshift observations of CO$(1\to0)$ emitters~\cite{2016ApJ...829...93K} and then extrapolated to high $z$. On the other hand, studies focused on the COMAP experiment~\cite{comap} model the CO(1-0) emission at $z\sim 2-3$ using COLDZ+GOODS measurements~\cite{2020ApJ...895...81R} as anchors. \par
In this work, however, we will adopt a data-driven approach to connect IR luminosity to CO luminosity for each transition and redshift. Using measurements of luminosity functions of $L'_{{\rm CO}(J\rightarrow J-1)}$ from galaxy-survey programs, we find values of $\alpha$ and $\beta$ which connect the $L_{\rm IR}$ distribution in our simulation to these data. In particular, for local galaxies we take results from Ref.~\cite{2016ApJ...829...93K}. For higher $J$ transitions at higher redshifts, we rely on results from the ASPECS survey using ALMA at 1.2 and 3 mm (250 and 100 GHz, respectively)~\cite{2020ApJ...902..110D}, and NOEMA results~\cite{Boogaard:2023whs} also at 1.2 and 3 mm bands. \par 
We use the measured CO luminosity functions at these 100 GHz and 250 GHz to anchor our theoretical predictions of CO contributions for a given CMB bandpass. Certain bandpasses such as 150 GHz do not have accompanying CO luminosity function observations, and so we will interpolate between 100 and 250 GHz observations when inferring $\alpha,\beta$ for that bandpass. We refer the reader to App.~\ref{appendix:interpolation} for more information.
\begin{figure*}[t!]
    \centering
    \includegraphics[width=0.7\textwidth]{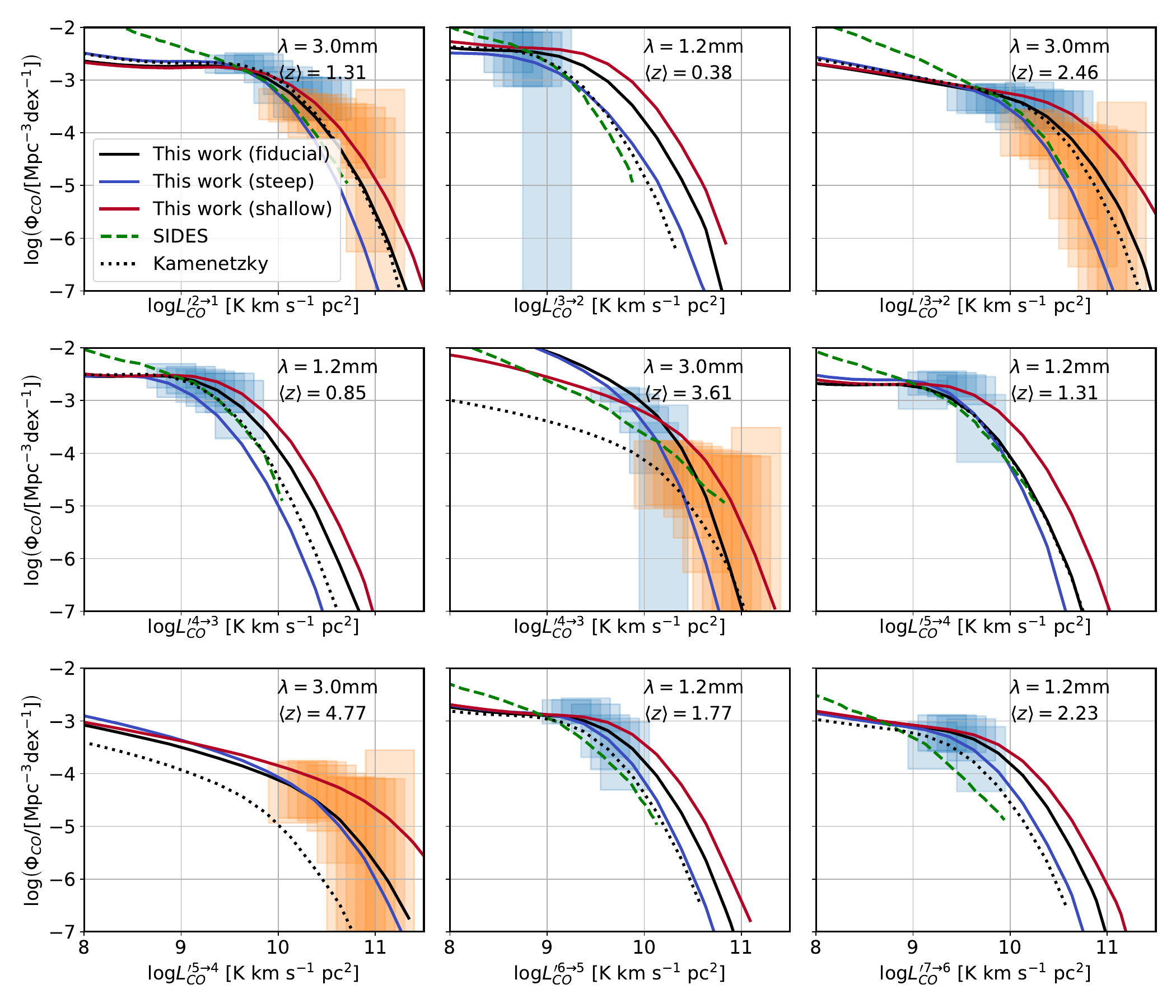}
    \caption{Comparison between the CO luminosity functions measured from sub-mm galaxy susrveys 
    (ASPECs, in blue squares and NOEMA in orange squares), and those obtained from the $\alpha-\beta$ parameterizations presented in this work. The steep, fiducial and shallow luminosity functions we adopt are shown in in blue, black and red, respectively. For each panel, we indicate the observed pivot wavelength of the measurements and the corresponding mean redshift of the sources observed. We also show, for comparison, the CO luminosity functions from \texttt{SIDES}~\cite{B_thermin_2022} (which underpins the analysis of Ref.~\cite{Maniyar_2023}, in dashed green lines), and the result of extrapolating the local measurements of the $L'_{\rm CO}-L_{\rm IR}$ relations from Kamenetzky~\cite{2016ApJ...829...93K} (in dotted black lines) to the redshifts considered in this work.}
    \label{fig:LFcollection}
\end{figure*}

We fit $\alpha-\beta$ values to the observed CO luminosity functions using the following procedure: 
\begin{enumerate}
    \item Given the frequency range of a luminosity-function measurement and the corresponding targeted CO transition $J$, we convert this to an observed range in redshift. 
    \item From the \agora halo lightcones, we select halos within that redshift range and compute their corresponding $\lir$ using the method discussed in Sec.~\ref{subsec:lir}. 
    \item Given an $\alpha-\beta$ set, we sample a corresponding $L'^{J\to J-1}_{CO}$ for every halo in the relevant lightcone using Eqn.~\ref{eqn:loglirco} and compute the resulting luminosity function of CO. 
    \item We iteratively sample over $\alpha,\beta$ using a simple least squares optimizer to converge on a reasonable value for the slope and amplitude. 
\end{enumerate}

We note that due to the paucity of data for the majority of these transitions,\footnote{For many radio luminosity functions there are at most two bins without overlapping $L'_{\rm CO}$ values.} especially at high luminosities for which 
it is exceedingly difficult that the small field of views surveyed contain such bright objects, there are many combinations of $\alpha_{\rm IR}$ and $\beta_{\rm IR}$ which all give acceptable fits.\par 
In order to quantify the impact of uncertainties in measured CO luminosity functions on the imprint of CO anisotropies in the CMB we generate, for each frequency band and transition, three separate pairs of $\alpha-\beta$ which are consistent with the data. We call these fits \emph{``steep'', ``fiducial''} and \emph{``shallow''} luminosity functions, as we find that the largest impact in varying $\alpha-\beta$ is at the high-$L_{\rm CO}$ end where the data are unconstrained.\footnote{We achieve this in practice by adding an artificial anchor point at $\log \Phi_{\rm CO} = -6$ and varying the value of $\log L'$ corresponding to this point. In principle one could also imagine jointly fitting $\alpha-\beta$ pairs and sampling from the resulting covariance matrix. However, given the lack of independent data for many of these bins or a suitable covariance matrix we have found that the fitting / sampling procedure is unstable. For bins where there are sufficient points for a stable sampling we find that the distributions of allowed CO LFs are similar to the range spanned by the steep/fiducial/shallow parameters presented.} We report these values in Table.~\ref{tab:alphabeta} and show the corresponding luminosity functions compared to data in Fig.~\ref{fig:LFcollection}. Note that Ref.~\cite{2016ApJ...829...93K} already provide the best-fit $\alpha_{\rm IR}$-$\beta_{\rm IR}$ values for the local luminosity functions; thus, we use these as our \textit{fiducial} case and do not consider steep or shallow scenarios. Note that, as we will see below, these transitions are very subdominant, so that this choice do not affect our results. We also compare our results, where available, to two previously adopted CO luminosity functions in the literature
\begin{itemize}
    \item The luminosity function from \texttt{SIDES}~\cite{B_thermin_2022} simulations, which were used to previously quantify the impact of CO as a contaminant to small-scale CMB anisotropies in \Maniyar. 
    \item Using the locally measured values of $\alpha_J$ and $\beta_J$ from low-redshift observations collected in Ref.~\cite{2016ApJ...829...93K} (henceforth referred to as the `Kamenetzky relations') and extrapolating them to all redshifts in our simulation. 
\end{itemize}
From Fig.~\ref{fig:LFcollection} we see that the ASPECs data primarily probes the ``knee" in the CO luminosity function, leaving significant uncertainty in the high-luminosity regime. Similarly, note that the luminosity functions derived from \texttt{SIDES} are comparable or steeper than our ``steep'' luminosity functions, where the `steepness' more accurately translates to the approximate exponential cut-off luminosity for a Schechter function fit to the data. Additionally, at lower luminosities than where data are available the \sides luminosity functions are significantly steeper than what is predicted from our $L_{\rm IR}$-based model. We note that varying the cutoff of our CO luminosity function at its bright end does not strongly alter its slope at the faint end -- it is stable for all transitions, including when we use the values from the Kamenetzky relations. \par 
The standard Kamenetzky relations, despite being built on an extrapolation of low-z information assuming that each transition has an identical power-law slope, seem to be in reasonable agreement with measured luminosity functions with the exception of the high-redshift $(5\to4)$ and $(4\to3)$ transitions. \par 

\subsection{Generating CO maps}
Given the above scheme to derive CO luminosities on a halo-by-halo basis from their underlying IR luminosities and a set of $\alpha-\beta$ values, we are now in a position to construct maps of the extragalactic CO that contributes to a given CMB bandpass. The adopted scheme is the same as detailed in Ref.~\cite{skyline}, but we summarize it below:
\begin{enumerate}
    \item Given a CMB frequency band, we identify the corresponding redshift range over which photons originating from a CO transition $J \to J-1$ will be redshifted into that band --shown in Table~\ref{tab:zs}.
    \item Shells of width $\Delta \chi = 50\,h^{-1}{\rm Mpc}$ that fall within this range are iteratively queried from the lightcone, and a CO luminosity is assigned to each halo using the derived IR luminosity (Eqn.~\ref{eqn:kennicutt}) and $\alpha-\beta$ values specified by the corresponding luminosity function, from Table.~\ref{tab:alphabeta}, following Eqns.~\ref{eqn:loglirco} and \ref{eqn:pseudoLconvert}.
    \item Every object in the shell is then given a Dirac delta ``SED"  { $\Phi_\nu = L_{\rm CO} \delta^{D}(\nu (1 + z_{\rm em}) - \nu_{\rm rest})$}, where $z_{\rm em}$ is the redshift of that halo which relates the rest-frame frequency to the observed through $\nu_{\rm obs} = \nu_{\rm rest}/(1 + z_{\rm em})$\footnote{{The Dirac delta function ``SED'' is defined in the rest frame of the object, and when integrating over the observed frequency bandpass this leads to an additional factor of $1/(1+z_{\rm em})$ in the observed flux. We thank Simon Foreman and Alex Van Engelen for pointing this out.}}.
    \item This ``SED'' is integrated against a spectral bandpass $\tau(\nu)$ for an experiment  in order to compute the flux at an angular position on the sky sourced by that object into that experiment. Since the SED is a delta-function {
    \begin{align}
    I_\nu^{\rm CO} &= \frac{L_{\rm CO} \tau(\nu_{\rm obs})}{4\pi \chi^2 (z_{\rm em}) (1 + z_{\rm em})^2 \int d\nu \tau(\nu) }, \\
    &\approx \frac{L_{\rm CO}}{4\pi \chi^2 (z_{\rm em}) (1 + z_{\rm em})^2 \Delta \nu}, \label{eqn:COflux}
    \end{align}
    }
    where we have assumed a top-hat pass band in the last equality to explicitly show the $\sim\Delta\nu$ suppression in the CO contribution to the measurements. We use actual ACT-like experimental bandpasses for the main results in this work, but we expect similar conclusions will hold for bandpasses of other ground-based experiments such as SPT or Simons Observatory.
    \item The flux is then converted into a spectral radiance by dividing $I_\nu / \Omega_{\rm pix}$, where $\Omega_{\rm pix}$ is the solid angle subtended by the desired HEALPix resolution.
    \item Given a bandpass we convert this spectral radiance into a CMB temperature decrement $\Delta T_\nu^{\rm CO}$, in units of $\mu$K, following Ref.~\cite{Planck_HFI}.
    \item The temperature decrement is summed to the map at the position of that object on the sky, and the previous steps are applied to a new shell, so that 
    $$ \Delta T_{\nu}^{{\rm CO}(J \to J-1)} (\btheta) = \sum_n^{N_{\rm shell}} \sum_{i \in {\rm shell\, n}} \Delta T_{\nu, i}^{{\rm CO}(J \to J-1)}(\btheta_i)\,. $$
    \item The final result is a HEALPix map which captures all CO$(J\to J-1)$ emission at a given frequency band, across all relevant redshifts. 
\end{enumerate}
This procedure produces a map of aggregate CO emission of all line emitters whose lines redshift into an experimental CMB bandpass, $\tau(\nu)$. We note again that, as the singular Dirac delta function SED gets integrated against the bandpass, this signal is suppressed by the width of the band $\Delta \nu$. \par 
In order to suppress bright outliers, every $J$ map also receives a rudimentary flux cut where the brightest $0.1\%$ pixels are masked out\footnote{This flux cut corresponds to $\sim 0.5$ mJy on average for each frequency band. However, the impact on spectra of using a 6.4 mJy cut such as the one in ref.~\cite{Reichardt_2021} is negligible on the CO spectra of our work other than the 220GHz 2$-$1 transition.}. This produces at most a 10\% change in the variance of the map, except for the 220 GHz CO$(2\to1)$ curve which probes the very local Universe, and we found that a small number of bright nearby sources fully dominated the spectrum of the map. 
\begin{figure*}[t!]
    \centering
    \includegraphics[width=1\linewidth]{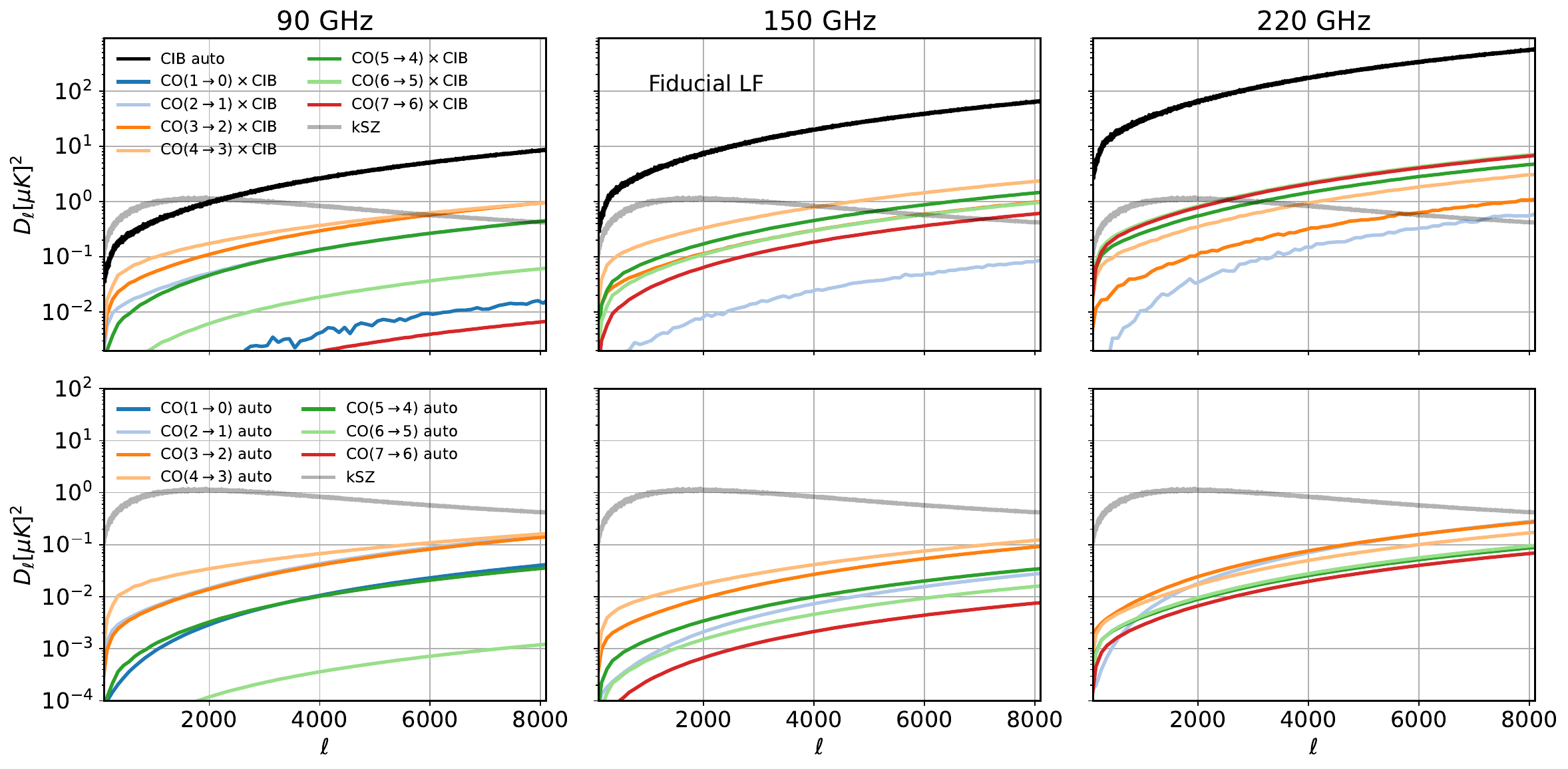}
    \caption{Auto and cross-spectra of individual CO transitions with CIB maps at 90, 150 and 220 GHz, for the ``Fiducial" luminosity function shown in Fig.~\ref{fig:LFcollection}, in comparison with CIB auto-spectra and the kSZ signal as predicted from \agora. }
    \label{fig:fidclustering}
\end{figure*}
\section{Results: angular power spectra of CO emission}
\label{sec:cells}
Having produced full-sky maps of CO emission for a transition $J\to J-1$, at a given bandpass $
\nu$, we measure their angular auto and cross power-spectra by using the standard \textsc{anafast} routine from the \texttt{HEALPix} library~\cite{Gorski_2005}. As the maps produced in this work cover the full sky, we do not require any additional treatment of mode-coupling induced by a survey mask. We additionally apply a $\ell_{\rm min} \sim 100$ for the largest angular scale considered, as the method used to construct lightcones from tiling the original \texttt{MDPL2} box does not include correlations on  scales larger than this. We refer to Ref.~\cite{skyline} for a more detailed discussion. \par 
In summary, we decompose the CO temperature perturbation maps into spherical harmonics
\begin{equation}
    \Delta T^{{\rm CO}(J\to J-1)}_{\nu} (\btheta) = \sum_{\ell, m=-\ell}^\ell a_{\ell m}^{{\rm CO}(J\to J-1),\, \nu} Y_{\ell m} (\btheta),
\end{equation}

and measure the corresponding auto and cross-spectra with maps of the CIB produced by \agora :
\begin{align}
    &C_\ell^{\rm CIB, {\rm CO}(J\to J-1)}(\nu) = \frac{1}{2\ell + 1} \sum_{m=-\ell}^\ell  a_{\ell m}^{{\rm CO}(J\to J-1),\, \nu} {a^*}_{\ell m}^{\rm CIB, \nu} \\
        &C_\ell^{\rm CO}(\nu) = \frac{1}{2\ell + 1} \sum_{m=-\ell}^\ell  \left | a_{\ell m}^{{\rm CO}(J\to J-1),\, \nu} \right |^2.
\end{align}
The resulting auto and cross-spectra for the fiducial luminosity functions adopted in this work are shown in Fig.~\ref{fig:fidclustering}. Specifically, we show the re-scaled power spectra
\begin{equation*}
    D_\ell \equiv \frac{\ell ( \ell + 1)}{2\pi} C_\ell,
\end{equation*}
in order to ease comparison with past work.
\subsection{The CIB $\times$ CO spectrum}
The upper panels of Fig.~\ref{fig:fidclustering} show the cross-correlation between CO emission and the CIB for three frequency bands corresponding to an ACT-like experiment, of 90,\footnote{The ACT 90 GHz band actually has an average frequency of 97GHz but we will refer to this band as 90 GHz throughout this text.} 150 and 220 GHz. Specifically, we show the full contribution to the observed TT power spectrum which is $2\times \langle {\rm CO} \times {\rm CIB} \rangle$, where $\langle {\rm CO} \times {\rm CIB} \rangle$ is the cross-spectrum between the CO map and the CIB map. The gray curve indicates the late-time kSZ signal from the \agora model with $\log( T_{\rm AGN} / {\rm K}) = 7.8$ and the solid black curves correspond to the CIB autocorrelation for the same bandpass in the fiducial \agora model. \par 
{In agreement with results reported in \Maniyar, we find that for 150 GHz the cross-spectra between aggregate CO emission and the CIB exceed $1 [\mu {\rm K}]^2$ at $\ell \sim 3000$. This amplitude is generically higher than the fiducial kSZ secondary anisotropies in \agora.} \par 
We now turn to a detailed breakdown of each CO transition's contribution to a given bandpass for the case of cross-correlations. It is helpful to consider the redshift ranges probed by each transition through each bandpass, remembering that the cosmic SFR density peaks at $z\sim2$, and that the average CIB intensity peaks at slightly higher redshifts. Figure~\ref{fig:freqs} contains a visual guide for several bandpasses and Table~\ref{tab:zs} specifies quantitatively the redshift probed by each frequency/transition combination assuming fiducial ACT-like bandpasses. \par 

Again referring to the upper panels of Fig.~\ref{fig:fidclustering}, we see that at 90 GHZ the amplitude of cross-correlation is primarily driven by the CO$(4\to3)$ transition with a subdominant contribution arising from $(3\to2)$. Past $\ell \gtrsim 3000$ the spectra for all transitions are similar to flat, Poisson-like shot noise. At this frequency, then, the lines that drive the CIB cross-correlation are located at $z\in [2.1-3.5]$ and $z\in[3.1-5.9]$.

At 150 GHz, the dominant transitions are CO$(4\to3)$ and CO$(5\to4)$, which following Table~\ref{tab:zs} correspond to emission from $z\in [1.7-2.7]$ and $z\in [2.4-3.6]$ -- again tracking the peak of cosmic SFR density around $z\sim2$ and the peak of CIB intensity at the slightly higher $z\sim 2.5$. Turning to the 220 GHz band the dominant transitions are CO$(7\to6)$ and CO$(6\to5)$ which are once again tracing the $z\sim 2-3$ region. \par 

  \begin{figure*}[t!]
    \centering
    \includegraphics[width=\textwidth]{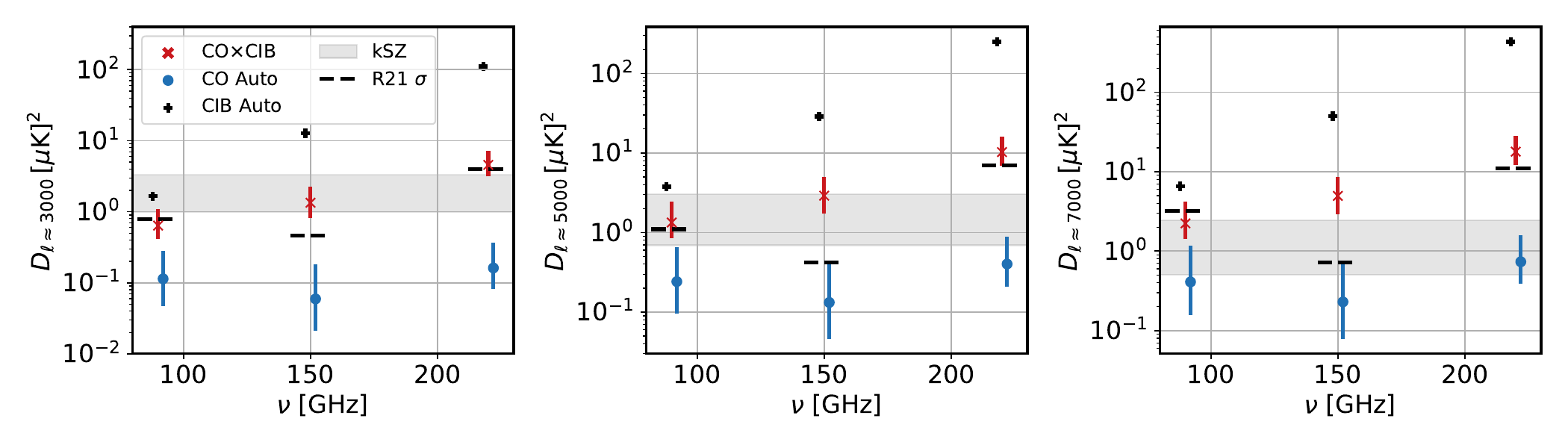}
    \caption{Aggregate contribution for all CO $J$ transitions considered in this work for the auto and cross-spectra, as a function of CMB band considered, for three different angular multipoles. The vertical bars associated with each contribution indicate the lower and upper bounds from assuming every $J$ transition follows its steep or shallow luminosity function, respectively. The gray shaded band corresponds to the kSZ amplitude from the three AGN temperature models discussed in the text. The dashed lines are the 1-$\sigma$ uncertainties from measurements at similar bandpasses for these angular multipoles, from a recent re-analysis of SPT data~\cite{Reichardt_2021}.}
    \label{fig:cosum}
\end{figure*}

Since the CIB is sourced by dusty star-forming galaxies and the CO abundance is similarly a tracer of star-formation and molecular gas, the abundance of which is correlated with that of dust, it is not surprising that the aggregate CO and the CIB show a high level of correlation and that the the prominent drivers of the CO$\times$CIB cross-correlation are consistently the lines which trace the epoch of peak star formation. \par 

Given the uncertainty on the luminosity functions of CO for the range of frequencies and transitions studied in this work (see Fig.~\ref{fig:LFcollection}) we study the impact of varying CO luminosity functions in the CO $\times$ CIB cross-correlation. This result is reported in Fig.~\ref{fig:cosum}, where the upper (lower) bars show the result of power spectra measured on maps generated with shallow (steep) luminosity functions. We find that the impact of uncertainty in the CO luminosity function (insofar as accounted for in this study) in this cross-correlation is appreciable but not sufficient to alter prior conclusions about the amplitude of this contribution relative to the kSZ effect. At $\ell=3000$ {we find
\begin{equation*} 
D_{\ell=3000}^{\rm CO \times CIB} \in \begin{cases}
    [ 0.4, 0.6, 1.1]\,\mu {\rm K}^2, \, \nu = 90 {\rm GHz}, \\
    [0.8, 1.3, 2.3]\, \mu {\rm K}^2, \, \nu = 150 {\rm GHz}, \\
   [3.1, 4.6, 7.2]\,  \mu {\rm K}^2, \, \nu = 220 {\rm GHz}, \\
\end{cases}
\end{equation*}}
 in the scenarios where all luminosity functions are simultaneously steep, fiducial or shallow, respectively. This corresponds to a factor of $\sim 2.3-2.6$ variation in the amplitude of this cross-correlation between the two extremes of scenarios entertained here.\footnote{These bands are to be taken as an estimate of the uncertainty arising from varying the luminosity function in the way described in \S~\ref{subsec:COLF}, but do not capture all possible uncertainty in CO emission given current known data.} In the fiducial \agora model, the three kSZ scenarios with $T_{\rm AGN} = \{10^{7.6},10^{7.8},10^{8.0}\} \, {\rm K} $ correspond to power spectra that span the range $D_{\ell = 3000}^{\rm KSZ} \in [1, 3.5] \mu{\rm K}^2$ across all frequencies. \par 

These findings corroborate the results shown in \Maniyar. The \sides luminosity functions are consistently comparable to the steepest luminosity functions consistent with sub-mm data, as well as having a faint luminosity slope that is steeper than what our model generically predicts. Additionally, we have shown that despite the strong scaling of the CIB amplitude with frequency, at 90 GHz this contaminant {can potentially be} comparable to the kSZ effect. Another conclusion one draws from the results in Fig.~\ref{fig:cosum} is that at 90 GHz the CIB$\times$CO cross-correlation is not only of a comparable amplitude to the kSZ effect, but also to the CIB itself. However, it is known that at these frequencies there are larger foregrounds such as the thermal Sunyaev-Zel'dovich effect and shot noise from radio sources which are individually on the order of $\sim 10 [\mu {\rm K}]^2$~\cite{Dunkley_2013}. Thus, we see that the CO$\times$CIB correlation is on the order of {4-11\%} of these components. Given the uncertainties at $\ell \sim 3000$ for recent ground-based experiments are on the order of $\sigma = 0.6 \mu {\rm K}^2$ this component {may be difficult to detect}.\footnote{Table 1 of Ref.~\cite{Reichardt_2021}}

\subsection{The CO $\times$ CO spectrum}
We now turn to the amplitude of the CO auto-spectrum and how it is impacted by uncertainties in CO luminosity functions. The CO auto-spectrum for each line is reported in the lower panels of Fig.~\ref{fig:fidclustering}. We find that the 90 and 150 GHz contributions are dominated by the CO($4\to3$) transition, and at higher amplitude than previously reported. This large amplitude can be understood as a result of the NOEMA measurements of CO$(4\to3)$ at 100 GHz, as shown in the central panel of Fig.~\ref{fig:LFcollection}. The NOEMA data prefer a distribution of CO luminosities at significantly higher values than those inferred solely from ASPECs data. Indeed, it has previously been noted that the luminosity function of this specific transition at 100 GHz is not well characterized by a single Schechter function~\cite{Boogaard:2023whs}. In contrast, the 220 GHz CO autospectrum receives roughly equal contributions from {the $(2\to 1)$, $(3\to 2)$, $(4\to 3)$ transitions}. \par 
While the CO auto-spectrum is suppressed by an additional factor of $1/\Delta \nu$ compared to the cross-correlation previously considered, our analysis highlights that this spectrum has a potentially non-negligible impact as a CMB extragalactic foreground when considered across the entire frequency range. Fig.~\ref{fig:cosum} highlights the aggregate contribution from CO auto as a CMB secondary, again in relation to the CIB, CIB$\times$CO and the kSZ signal. {We find that, in agreement with priors works, the CO auto contribution is sub-dominant compared to other components considered in this work and outside of the range of detectability for an SPT-like experiment.} Across the three frequency bands considered we have {
\begin{equation*}
D_{\ell=3000}^{\rm CO \times CO } \in 
    \begin{cases}
        [ 0.05, 0.11, 0.28]\,\mu {\rm K}^2,\,\nu = 90 {\rm GHz}\\
        [0.02, 0.06, 0.2]]\,\mu {\rm K}^2,\,\nu = 150 {\rm GHz}\\
        [0.08, 0.16, 0.36]\, \mu {\rm K}^2,\,\nu = 220 {\rm GHz}
    \end{cases}
\end{equation*}}
This observed -- nearly six-fold --variation in the amplitude of CO auto at small scales from uncertainties in the CO luminosity function is significantly larger than what we observed in cross-correlations. This is expected, since the uncertainties are compounded in the case of the auto spectra. The CO auto contribution is suppressed at higher frequencies compared to CO$\times$CIB, {and for an experiment with error bars comparable to Ref.~\cite{Reichardt_2021} it will probably not be a detectable component.} 
\begin{figure*}
    \centering
    \includegraphics[width=\textwidth]{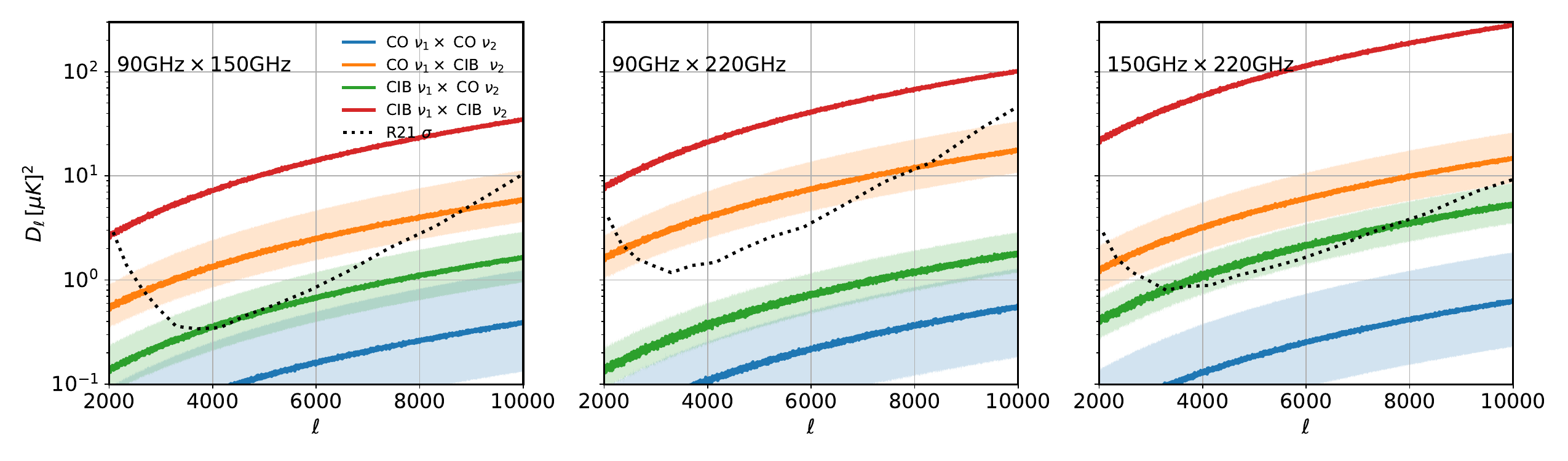}
    \caption{Aggregate CO and CIB auto and cross-correlations between different frequency channels, for transitions up to $J=7$. The dotted black line corresponds to the 1-$\sigma$ of measurements from SPT for similar bandpasses, from Ref.~\cite{Reichardt_2021}.}
    \label{fig:crossnu}
\end{figure*}
\subsection{Cross-frequency spectra}
Having characterized the frequency dependence of the auto and cross-spectra of CO emission in ground-based CMB experiments, we now examine the amplitude of power spectra at differing frequency channels. The cross-frequency spectrum is a complementary probe of CMB extragalactic foregrounds:  signals which have a large autocorrelation in one channel can have significantly lowered cross-correlation with another channel, since components such as radio galaxy counts or CIB intensity scale with frequency but in an opposite fashion. In addition, detector-specific noise, which can be prominent in auto-correlations, is not present in cross-correlations. \par 
We measure these cross-frequency spectra with aggregate maps of CO emission and CIB emission for the three bandpasses considered in this work. These are shown in Fig.~\ref{fig:crossnu}. We observe the CIB cross-frequency spectrum is always dominant relative to any correlation involving just CO lines, {and the low-high CO$\times$CIB cross-correlation (where CO emission comes from the lower frequency map and CIB from the higher frequency map) is the second highest component}. Specifically, we find for our fiducial model that {
\begin{equation*} 
D_{\ell=3000} \in \begin{cases}
    [ 4.5,\,13.2]\,\mu {\rm K}^2, \, {\rm CIB_1\times CIB_2}, \\
    [0.9,\, 2.6]\, \mu {\rm K}^2, \, {\rm CO_1\times CIB_2},
\end{cases}
\end{equation*}}
for the $90\times 150$ GHz and $90\times 220$ GHz cross correlations, respectively. 
In the SPT analysis presented in Ref.~\cite{Reichardt_2021}, the dominant contribution to cross-frequency spectra involving the 90 GHz band was CIB contamination. Given that our fiducial model for CO emission at 90 GHz correlates with the CIB to {be an appreciable fraction of the CIB auto}, we conclude that the CO cross-correlation is particularly important in cross-frequency analyses. \par 
The cross-frequency spectrum is particularly interesting if we also consider how uncertainties in CO luminosity functions impact this cross-correlation. The bands in Fig.~\ref{fig:crossnu} denote the variation from the ``steep'' to the ``shallow'' luminosity functions. An analysis of cross-frequency small-scale CMB secondaries that includes the templates of CO emission should be particularly sensitive to this amplitude. As a heuristic guide for how sensitive current (and future) cross-frequency measurements will be to these components, we also plot the 1-$\sigma$ uncertainties measured from Ref.~\cite{Reichardt_2021} for similar bandpasses as those considered in this work. We see that the cross-frequency spectra are measured at exquisite precision, and even components such as the {CIB$_1 \times$CO$_2$ correlation for $90 \times 150$ GHz appear to be comparable to measured uncertainties}. Given the relevance of cross-frequency correlations for template-based analyses and component-separation techniques, we conclude that contributions from CO should be considered to avoid suboptimal or even biased results. \par 
The amplitude of this cross-correlation is driven by the $J=4\to 3$ transition and in particular the high-$L$ tail measured by Ref.~\cite{Boogaard:2023whs} -- around 45\% of the total ${\rm CO}_1 \times {\rm CIB}_2$ amplitude for $\nu_1 = 90,\,\nu_2=150$ GHz comes from this particular transition.

\section{Discussion}
\label{sec:discussion}
\subsection{Analytic scalings for the shot-noise term}
We have generated maps of extragalactic CO emission at various redshifts conditioned on observations of CO luminosity functions from small-field-of-view sub-mm surveys at similar frequencies. The measurements of the luminosity functions employed allow for freedom in the predicted high-$L'$ tail, which in turn results in a significant uncertainty in the amplitude of CO auto correlations and its cross correlations with other CMB secondary anisotropies and extragalactic foregrounds. 
Here we discuss a simple analytic argument which connects the amplitude of the spectra to the luminosity function, focusing in particular on the uncertainties at the bright end of the luminosity functions.\par 
In Appendix~\ref{appendix:poisson} we show that for the majority of cross-spectra measured in this work, the contribution at $\ell \sim 3000$ (the best-measured multipole) becomes driven by the shot-noise contribution, rather than the clustering component. For tracer auto-spectra, it is well-known that 
the shot-noise contribution essentially corresponds to  the second moment of the flux distribution\footnote{Note that this is only exact for Poissonian shot noise. However, nonlinear effects such as halo exclusion and beyond-linear galaxy bias introduce deviations from this predictions; see e.g., Refs.~\cite{Moradinezhad_Dizgah_2022b,Obuljen:2022cjo} for examples in the context of line-intensity mapping.}~\cite{Dunkley_2013}
\begin{equation}
    C_{\ell}^{\rm shot} =  \int dS \frac{dN}{dS} S^2,
\end{equation}
where $N$ is the absolute number of emitters in the volume probed by that frequency band. For the CO contribution of a given $J$ transition, we can condition the flux distribution on redshift through the bandpass. Assuming for this argument a top-hat bandpass for simplicity, we are left with
\begin{align}
        C_{\ell}^{\rm shot} &=  \int dS \frac{dN}{dS} S^2 \\
        &= \int_{z_{\rm min}}^{z_{\rm max}} dz\,dS \frac{d^2 N}{dz dS} S^2
\end{align}
where $[z_{\rm min}, z_{\rm max}] = [\nu_{\rm rest}/\nu_{\rm max} - 1, \nu_{\rm rest}/\nu_{\rm min} - 1]$. 
Converting fluxes to luminosities (using Eqn.~\ref{eqn:COflux}) we get
\begin{align}
        C_{\ell}^{\rm shot} &= \int_{z_{\rm min}}^{z_{\rm max}} dz\,dS \frac{d^2 N}{dz dS} S^2 \nonumber\\
         &= \int_{z_{\rm min}}^{z_{\rm max}} dz\,dL \frac{d^2 N}{dz dL} \frac{1}{\Delta \nu ^2} \left ( \frac{L}{4\pi (1+z)\chi^2(z)} \right )^2 \nonumber\\
        &= \frac{c}{4\pi (\Delta \nu)^2} \int_{z_{\rm min}}^{z_{\rm max}} \frac{dz}{H(z)}\,dL\, \frac{d^2 n}{dz dL} \frac{L^2}{\chi^2(z) (1+z)^2}\,, \label{eqn:shotnoise}
\end{align}
where $n$ is now, instead, the number density of emitters. Eqn.~\ref{eqn:shotnoise} then makes it clear how a shallower luminosity function at a given redshift will fundamentally lead to a higher shot noise. Eqn.~\ref{eqn:shotnoise} also shows that what is fundamentally being probed is the joint redshift--luminosity distribution function of line emitters in the Universe. \par 
If one wishes to extend this discussion to cross-line emission or CO$\times$CIB correlations, then one must realize that the underlying driver of CO luminosity is the IR luminosity (Eqn.~\ref{eqn:lcolir}), or SFR/$M_*$ (Eqn.~\ref{eqn:kennicutt}) and the shot-noise is then driven by expectations of the sort 
\begin{equation*}
C_\ell^{\rm shot} \sim \langle L_{\rm CO}(L_{\rm IR}) \times L_{\rm IR} \rangle\,\, {\rm or }\,\, \langle L_{\rm CO}({\rm SFR}) \times L_{\rm IR}({\rm SFR})\rangle.
\end{equation*}
These correlations are equivalent to Eqn.~\ref{eqn:shotnoise}, but changing the $L^2(d^2n/dzdL)$ term by a term like $\propto L_{\rm CO}L_{\rm IR}(d^3n/dzdL_{\rm CO}dL_{\rm IR})$, with the corresponding correlation between the CO and IR luminosities as discussed in Appendix~\ref{appendix:cross-shotnoise}.\par 
For models such as Eqn.~\ref{eqn:lcolir} where we have $L_{\rm CO} \propto L_{\rm IR}^{1/\alpha_{J,{\rm IR}}}$ then we see that the CO$\times$CIB does \emph{not} probe the second moment of the $L_{\rm IR}$ distribution, but really in some sense the expectation $\langle L_{\rm IR}^{1 + 1/\alpha} \rangle$ -- we refer the reader to Appendix~\ref{appendix:cross-shotnoise} for a more detailed discussion of the formulae of cross shot noise and under what conditions the above expectation is in fact being probed. \par 
Let us consider now a toy model where all emission comes from a delta-function distribution of redshift to get additional insight on the relation between the luminosity function and the Poisson shot noise. In this toy model the geometric factors are fixed and do not contribute to changes in the variance of fluxes, and changes in the luminosity functions are the only sources of changes to the shot noise. We note that for many of the panels in Fig.~\ref{fig:LFcollection}, a Schechter function of the form
\begin{equation}
    dn = n_* \left ( \frac{L}{L_*}  \right )^\alpha  e^{-L/L_*} d\left(\frac{L}{L_*}\right),
\end{equation}
can fit the model curves, despite not being an input into either \sides or our fiducial model for CO luminosities. For concreteness we take the luminosity functions for CO$(2\to1)$ at 90 GHz and fit $n_*, \alpha, L_*$ to the reported curves. We find that $\alpha = -0.6$, $L_* = 1.2 \times 10^{10} L_\odot$ and $\alpha = -0.2$, $L_* = 1.5 \times 10^{10} L_\odot$ are a reasonable fit to the luminosity functions of \sides and the fiducial \skyline scenario, respectively. The inferred values of $n_*$ are very similar for both cases. This is is in agreement with the requirement of recovering the total number density of CO emitters in the redshift shell. Assuming there is a minimal cutoff CO luminosity $L_0$ 
below which halos do not 
emit CO,  as well as extending the upper limit of the integral to infinity, the variance of CO luminosity is given by (up to a volumetric factor) 
\begin{figure}
    \centering
    \includegraphics[width=1\linewidth]{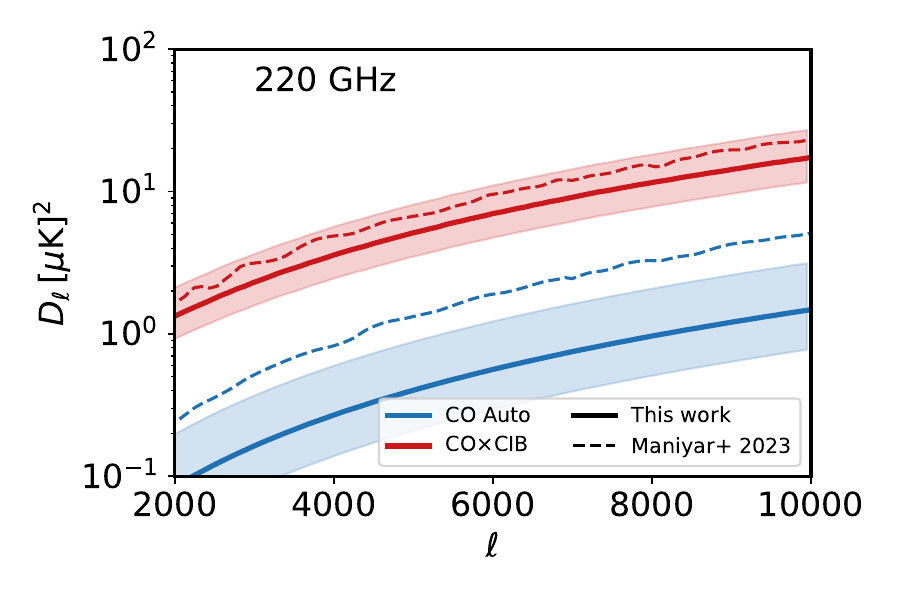}
    \caption{{Templates for aggregate CO contamination at 220 GHz compared to the predictions from the \sides simulation as shown in M23. The CO $\times$ CIB cross correlation is consistent between both simulations, however we see that our simulations predict a lower CO auto-spectrum than M23. In this figure we explicitly show the correlation $\langle {\rm CO} \times {\rm CIB} \rangle$ instead of its aggregate contribution to CMB spectra as shown in prior figures. The shaded bands around our fiducial predictions encompass the steep and shallow luminosity functions considered in this work.}}
    \label{fig:220 GHzsidescompare}
\end{figure}
\begin{align}
\langle L^2 \rangle &\propto \int \frac{dn}{dL} L^2 dL \\
&\propto {L_*}^2 \int_{L_0}^\infty \left ( \frac{L}{L_*}  \right )^{2+\alpha}  e^{-L/L_*}\, d(L/L_*) \\
&\propto n_* {L_*}^2 \Gamma(3+ \alpha, \frac{L_0}{L_*}),  
\end{align}
where $\Gamma(3+\alpha, \frac{L_0}{L_*})$ is the upper incomplete Gamma function~\cite{NIST:DLMF}. The incomplete Gamma function is a strictly monotonically decreasing function in its second argument, peaking at $L_0 / L_* = 0$, while monotonically increasing in its first argument, $3+\alpha$. Therefore, we find that both a higher $L_*$ at fixed $\alpha$ and a higher $\alpha$ at fixed $L_*$ lead to a higher $\langle L^2\rangle$. Since for \sides $\alpha=-0.6$, which is lower than the value for \skyline, and their luminosity functions frequently cut off at $L_*$ consistent with our `steep' luminosity functions -- corresponding to low values of $L_*$ --, we should generically expect that our model will produce CO maps with higher shot noise components than those of \sides. Assuming $\log L_0 = 8$, we find the second argument of the upper incomplete Gamma function is close to saturated and we may neglect its dependence. Then, for the specific parameters in this scenario we find that the ratio of moments is 
\begin{equation*} 
\frac{\langle L^2  \rangle_{ \rm \tiny SkyLine}}{\langle L^2  \rangle_{\rm \tiny SIDES}}  \sim \left ( \frac{1.5}{1.2} \right)^2 \frac{\Gamma(2.8)}{\Gamma(2.4)} \sim 2.1,
\end{equation*}
 which are of similar order to the differences in CO shot noise we will find in the next section. \par 
Since the joint IR-CO shot noise will be driven by similar underlying mechanisms we expect that these arguments should similarly hold. However, it would be important to examine the underlying distribution of IR fluxes / luminosities between the two simulations and see if the same qualitative differences are observed compared to the CO luminosity functions we have examined in this work. Another challenge in the case of IR luminosity is that the broad SEDs require some additional care compared to our treatment of CO line emission which can be readily treated as a Delta function -- again we refer to Appendix~\ref{appendix:cross-shotnoise} for more discussion on this note. 

\begin{figure}
        \centering
        \includegraphics[width=\linewidth]{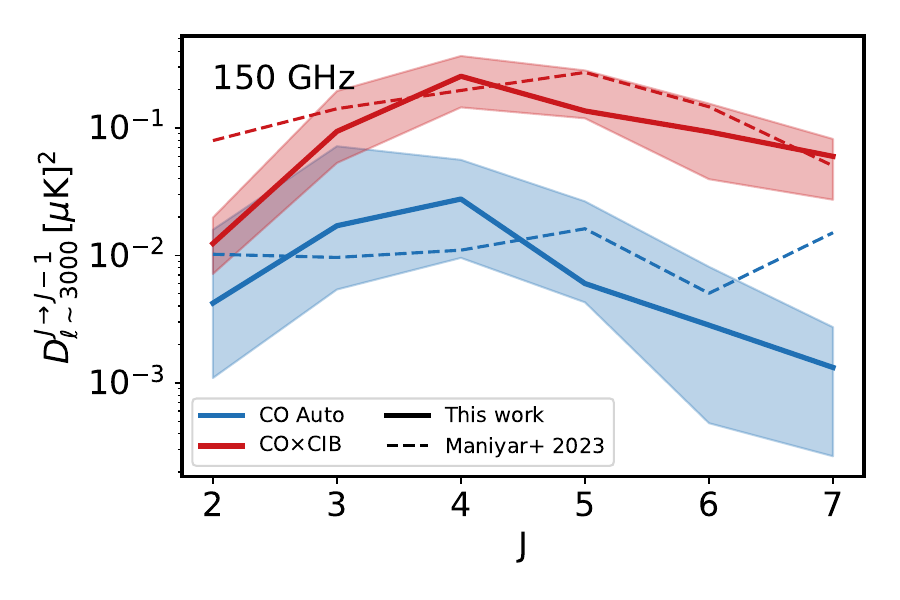}
        \caption{The amplitude of CO power spectra at $\ell=3000$ and 150 GHz for each transition from CO$(2-1)$ to CO$(7-6)$ compared to the \sides predictions reported in M23.}
        \label{fig:150 GHzJcomparesides}
    \end{figure}
\subsection{Comparison with previous work}
\begin{figure*}
    \centering
    \includegraphics[width=0.9\linewidth]{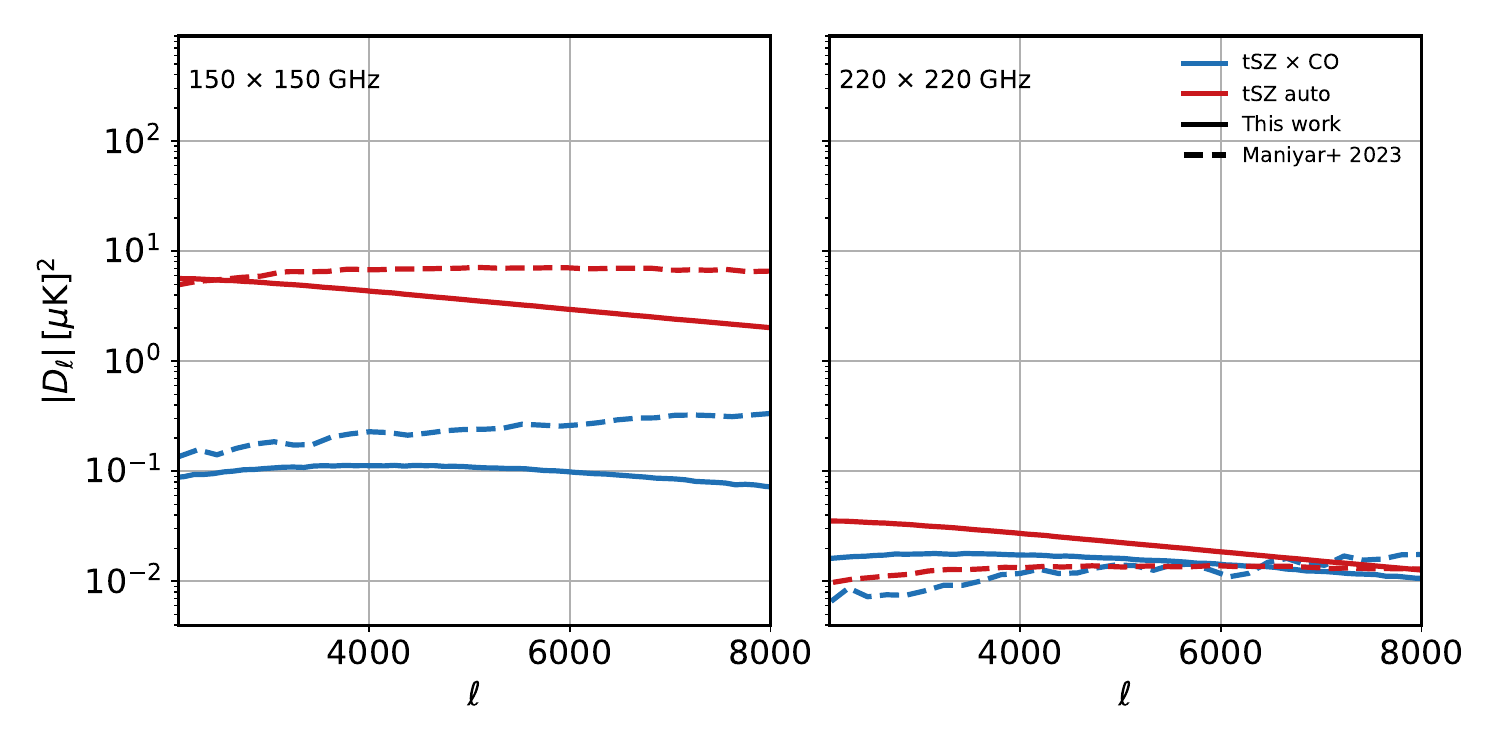}
    \caption{Cross-correlations between tSZ maps and CO maps from our templates, in comparison to the \sides results from M23. The cross-correlations are overall similar, but we note that both models predict different tSZ signatures.}
    \label{fig:tszco}
\end{figure*}
There has been previous work to characterize the amplitude of extragalactic CO signal and its cross-correlation with other sources of anisotropies such as the CIB and the tSZ, mainly the work presented in \Maniyar. In this section we proceed to compare when possible the predictions generated from our model with theirs. Mainly, we compare to the published aggregate CO autospectra and CO$\times$CIB spectra for a 220 GHz bandpass, as well as a line-by-line comparison for the auto and cross contributions at 150 GHz. We caution that these comparisons will not be one-to-one for many reasons, including that we have used the full ACT bandpass in this work whereas the study presented in \Maniyar uses top-hat bandpasses of width $\Delta \nu = 50$GHz, as well as adopting a different treatment to mask bright sources. The predictions shown in \Maniyar are predicated on the CONCERTO-\sides simulation~\cite{B_thermin_2022}, which were developed with the intention of mocking up [CII] intensity mapping surveys at high redshift for which extragalactic CO is a significant contaminant. The underlying mechanism for
modeling the luminosity of CO lines is also entirely different from the empirical approach we have adopted. 
However, it is still interesting to see how these different approaches to simulating the same signal compare to each other. \par 
In Fig.~\ref{fig:220 GHzsidescompare} we show the aggregate CO auto and CO$\times$CIB spectra for the 220 GHz band as generated in this work and reported in \Maniyar. {We see that at this frequency, our fiducial scenario generates a CO$\times$CIB spectrum that is consistent with their work. However, we also see that our fiducial scenario generates a prediction for the CO auto-spectrum which has similar angular dependence (consistent with being shot noise) but has a lower amplitude than the prediction from \sides. The \sides prediction is $\sim$ 50\% higher than our shallow scenario, and is close to thrice higher than our fiducial scenario. As discussed previously, and seen in Fig.~\ref{fig:fidclustering}, our 220GHZ CO auto-spectrum is driven by the nearest transitions -- $(3\to 2)$ and $(2\to 1)$, while the cross-spectrum is driven by $(7\to 6)$ and $(6\to5)$, whose contributions to the band are entirely contained within the peak redshift of the CIB intensity. However, we do not have similar line-by-line comparisons of the \sides 220 GHz spectrum to assess which are their dominant contributions.} \par 
For 150 GHz, however, \Maniyar breaks down their CO auto and cross spectra by transition. Therefore, in Fig.~\ref{fig:150 GHzJcomparesides} we have elected to compare our predictions at the fixed angular scale of $\ell=3000$ in order to compare how each $J$ line contributes to the observed band in each model. {We find that at 150 GHz our model is similar in amplitude for both auto and cross-spectra. For cross-spectra there is some broad agreement, and the $J$ dependence has a similar `rising and falling' shape across both models, with a peak at $J=4$ for our model and $J=5$ for \sides. The auto-spectra are different in amplitude and $J$-dependence. Our model predicts, again, a peak at $J=4$ with $D_{\ell\sim 3000}^{4\to 3} \sim 2 \times 10^{-2} [\mu {\rm K}]^2$ and a faster fall-off than the CIB cross-correlation. For \sides, however, the J dependence is nearly flat and $D_{\ell\sim 3000} \sim 10^{-2} [\mu {\rm K}]^2$ for nearly all of their spectra.} It would be especially interesting to compare the predictions of CO luminosity functions between these two models at 150 GHz, as there are no surveys trying to measure this observable at these wavelengths. However, we leave this detailed comparison between these two models for future work. \par 

{The analytic scaling arguments presented in the previous sub-section can qualitatively explain why the fiducial model in this work generically predicts somewhat different amplitudes of CO auto and cross-spectra with the CIB compared to previous attempts to characterize this signal. In this subsequent subsection we have further investigated and quantified differences between the two models, focusing on 150 and 220 GHz where the CONCERTO-\sides suite has been used to generate predictions for extragalactic CO contamination. For 220 GHz we find a strong enhancement of the CO auto relative to our results, and somewhat comparable CO$\times$CIB spectra. For 150 GHz both the CO auto and cross-spectra are within the bands from our range of scenarios, with the exception of the $(2\to1)$ cross-spectrum and the $(7\to6)$ CO auto-spectrum, where \sides is above even our shallow luminosity functions.} \par 
Finally, we also also present a comparison of the the tSZ $\times$ CO correlation predicted by our model, using \agora tSZ maps, in Fig.~\ref{fig:tszco}, along with the results from \Maniyar. Both models return similar predictions for the cross correlation. While the signal is very small -- halos that host an appreciable tSZ signal are at low redshift and very massive, and do not simultaneously host significant CO luminosity, since their star formation is low -- this test is interesting as it probes CO emission models in very different halo mass regimes than the auto-spectra or cross-correlations with the CIB. {We find that at 150 and 220 GHz the tSZ$\times$CO signal is comparable in amplitude between both models. Differences in the $\ell$-dependences and amplitudes for both frequency channels are consistent with differences in the observed tSZ auto-spectra.}

\subsection{Comments on the predicted luminosity functions}

As discussed in Sec.~\ref{sec:sims}, we use $\alpha_{\rm IR}$-$\beta_{\rm IR}$ values obtained from fits to measured luminosity functions, with the corresponding uncertainties and steep/fiducial/shallow scenarios. However, as shown in Fig.~\ref{fig:LFcollection}, in all cases but the $(4\to3)$ transition at $\langle z\rangle=3.61$, the luminosity functions produced for all three scenarios feature a very similar faint end. This might be due to limitations on the parameterization based on Eqn.~\eqref{eqn:loglirco} but also due to the limited resolution of the MDPL2 simulations employed as underlying halo distribution.\footnote{However, see Appendix A of \cite{skyline} where it was shown that at $z=5$ around 90\% of the total cosmic SFR density is reproduced due to resolution limitations, and so the fact the faint-end slope in our simulations remains constant down to low redshifts points against the resolution being an issue. } In turn, \sides used the Uchuu simulations, which has a higher resolution, which may explain their higher faint end of the luminosity function. In any case, estimations of the shot-noise contribution 
 -- which dominates the signal of interest -- using the Schechter fit discussed above show that this difference amounts to $<1\%$ difference in the predicted power spectra. Nonetheless, this feature warrants further investigation, given its relevance for line-intensity mapping dedicated surveys, which target the faint end of the luminosity functions.\par 
 
We also note, in passing, that another fundamental driver in discrepancies could be the IR luminosity modeling of both approaches. Indeed, the \sides predictions tend to somewhat under-predict the amplitude of the CIB power spectra compared to Planck and Herschel/SPIRE data (see Fig. 3 of Ref.~\cite{Gkogkou:2022bzo}) and we would expect that under-estimating IR anisotropies could lead to compounding differences between our two predictions for CO emission. Their usage of a pure Kennicutt relation to assign IR luminosities from star formation rates would also lead to significant changes in the faint-end of the $L_{\rm IR}$ distribution, which may result in the higher abundance of faint CO emitters predicted with \sides with respect to our results. It would be of great interest to further explore differences in the $L_{\rm IR}$ distributions between different simulations of extragalactic foregrounds, since these are the initial building blocks of CIB and IR line-emission models.
\section{Conclusions}
\label{sec:conclusions}
In this work we have used the semi-empirical intensity mapping modeling code, \skyline, conditioned on observations of CO luminosity functions, from $J=1$ to $J=7$, covering $z\sim 0$ to $z\sim 9$, to generate maps of extragalactic CO emission through bandpasses of ground-based low frequency CMB surveys, specifically for ACT-like bandpasses at 90, 150, and 220 GHz. We generated three sets of maps, corresponding to CO luminosity functions featuring bright-end cutoffs at different luminosities, as the available data do not always probe this regime with enough precision. Our fiducial model often generates a CO luminosity function very similar to assuming locally measured scaling relations from Ref.~\cite{2016ApJ...829...93K}, except for the high-redshift $(4\to3)$ and $(5\to4)$ transitions measured at 100 GHz. 

We have used these maps to measure each line's power spectra and cross-spectra with accurate simulated maps of the CIB from the \agora simulations, whose large-scale structure, star formation, and derived IR luminosities are self-consistent with our simulation. We have found, consistent with prior expectations, that the aggregate CO$\times$CIB spectrum is always larger in amplitude than the CO auto-spectrum, with an angular dependence for both types of spectra that is wholly consistent with the shot-noise component at $\ell \gtrsim 2000$, for all frequencies considered. We additionally found that the aggregate signal, at $\ell \sim 3000$, is comparable to or brighter than the late-time kSZ power spectrum from the the \agora simulation.  \par 
For the first time we also investigated the cross-frequency spectrum of extragalactic CO, finding that cross-correlations of CO emission at 90 GHz with the CIB at 150 GHz {were not negligible compared to the 1$\sigma$ uncertainties of this cross-frequency spectrum}. Given that this cross-frequency spectrum is exquisitely well measured, and that the dominant extragalactic foreground considered thus far has been this CIB cross-spectrum, these results indicate a promising avenue through which this extragalactic CO foreground can be constrained. This ${\rm CO}_1 \times {\rm CIB}_2$ spectrum may be the second-largest foreground behind ${\rm CIB}_1 \times {\rm CIB}_2$ at 90 GHz $\times$ 220 GHz. \par 
A brief investigation was also conducted comparing the predictions of our empirical model to the CONCERTO-\sides suite of 
simulations~\cite{B_thermin_2022}, which were used in \Maniyar to similarly quantify this extragalactic CO foreground. {We found broad agreement at 150 GHz but disagreement in our results, especially the amplitude of CO auto spectra, at 220 GHz.} Given the potential impact of including templates of CO emission motivated by our simulations in analyses of CMB survey data a more detailed comparison between the two models is of pressing importance in follow-up studies. \par 
The measured spectra can be readily folded into analyses of ground-based CMB data, and the tools we have developed can be extended to any bandpass for any existing CMB survey. We believe it will be interesting to see what the impact of including our templates is, for example, in inference on the amplitude of the kSZ effect in the CMB temperature power spectrum, or even in potential inference on the shot-noise component of the CIB which we expect is highly correlated with the CO signal. Furthermore, neglecting the extragalactic CO foreground could impact component-separation analyses of the CMB. [CII] emission ($\nu_{\rm rest}=1.9$ THz) might also be expected to be a contaminant for the \textit{Planck} HFI bands. \par 
The maps in this work\footnote{The $C_\ell$s and notebooks used to recreate figures in this work will be made publicly available in \href{https://github.com/kokron/skyLine/tree/main/examples/}{https://github.com/kokron/skyLine/tree/main/examples}. The maps are available upon reasonable request.} can also be used to readily investigate other potential sources of contamination by CO in CMB cross-correlations. For example, an analysis of galaxies from the unWISE galaxy catalog (which are IR luminous, and therefore should also have CO emission) revealed a detection of a higher than expected amplitude for the kSZ effect~\cite{Kusiak:2021hai}. It is not inconceivable that correlations between unWISE galaxies and unresolved CO emission in a CMB map could be detected at high significance. Indeed, a detection of diffuse [CII] emission at $z\sim 2.6$ was reported using a cross correlation of Planck HFI and BOSS quasars~\cite{Yang:2019eoj}. Other potentially interesting avenues of our work could be related to using higher-order statistics of the CMB temperature map or in cross correlation with tracers of the large-scale structure, such as bispectra, to tease out components which allow us to better constrain the properties of CO emission through a given bandpass and ideally break degeneracies between the emission from different transitions. If successful, these investigations would be complementary to line-intensity mapping experiments, which are expected to probe the faint-end of the line-luminosity functions. \par
Another possible scenario is that the low-frequency Poisson components of current CIB models compensate for the presence of CO emission. For example, the \agora CIB model has a Poisson component that fully accounts for the small-scale cross-frequency power spectrum (such as 90 $\times$ 150 GHz). The addition of a CO $\times$ CIB component at the fiducial values here could lead to a potential over-estimate of small-scale power for this frequency correlation. The correct procedure, then, would be to jointly account for the contributions of CO and CIB when fitting to small-scale spectra across a wide range of frequency bands. Since CO luminosity is sourced by IR luminosity, we expect these amplitudes to be correlated. Lowering the amplitude of the CIB component (since we have properly accounted for its joint presence with CO) would also potentially lead to lower amplitude CO components. We defer these sorts of investigations to future work. \par 
We conclude with a final remark that it would be highly desirable to develop an accurate and simply parameterized model for the CO emission that could be easily folded into small-scale CMB likelihood analyses. Our fiducial simulations show that there is significant uncertainty in the amplitude of this signal, and a compact parameterization that properly captures both auto and cross-clustering across frequencies, would be highly desirable. For example, a parameterized conditional distribution of IR and CO luminosity that can be used as input into Eqn.~\ref{eqn:shotnoise}. This is not expected to be trivial -- each bandpass probes each transition at significantly different cosmic epochs, and even the CIB probes different epochs at different bandpasses. It is not expected that a simple frequency-scaling exists that can capture such large time evolution, but we leave these investigations and this construction to future work. In light of the significant improvements expected to come in the next generation of ground-based observations, such as those from Simons Observatory, we believe developing a model for such a potentially prominent astrophysical signal should be significantly timely.

\begin{acknowledgments}
We thank Raul Abramo, J. Colin Hill, Louis Legrand, Abhishek Maniyar, and Gabriela Sato-Polito for enlightening discussions. We also thank Yuuki Omori for making unlensed \agora CIB maps that were used in this work publicly available. NK and JD acknowledge support from NSF award AST-2108126. NK would like to thank the International Center for Theoretical Physics – South American Institute for Fundamental Research (ICTP-SAIFR) for their hospitality during part of the completion of this work. ICTP-SAIFR is funded by FAPESP grant 2021/14335-0. JLB acknowledges funding from the Ramón y Cajal Grant RYC2021-033191-I, financed by MCIN/AEI/10.13039/501100011033 and by
the European Union “NextGenerationEU”/PRTR, as well as the project UC-LIME (PID2022-140670NA-I00), financed by MCIN/AEI/ 10.13039/501100011033/FEDER, UE. Figures and code to produce them in this work have been made using the SciPy Stack \citep{2020NumPy-Array,2020SciPy-NMeth,4160265}. This research has made use of NASA's Astrophysics Data System and the arXiv preprint server. \par
Some of the computing for this project was performed on the Sherlock cluster at Stanford. We would like to thank Stanford University and the Stanford Research Computing Center for providing computational resources and support that contributed to these research results.
\end{acknowledgments}

\appendix

 \onecolumngrid

\section{Choice of $\alpha_{\rm IR}$-$\beta_{\rm IR}$ values for transitions and redshifts not covered by observations}
\label{appendix:interpolation}
As described in Sec.~\ref{subsec:COLF}, we follow a data-driven approach to model the CO emission at each transition and redshift we need. However, not all cases of interest are probed by observations. This is the case for the $(6\to5)$ and $(7\to6)$ transitions at 90 GHz, and all transitions at 150 GHz. In order to fill this gap, we use the existing measurements for other transitions and frequencies to inform the choice of suitable $\alpha_{\rm IR}$-$\beta_{\rm IR}$ sets, following two different approaches.

\begin{figure}
     \centering
     \includegraphics[width=0.5\columnwidth]{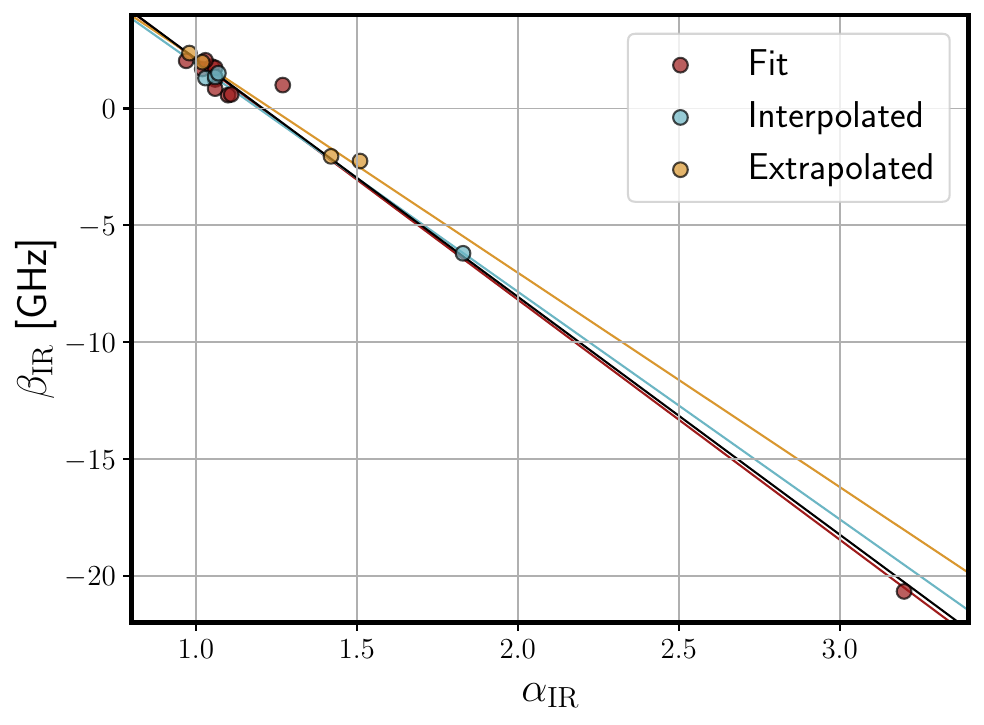}
     \caption{Correlation between the $\alpha_{\rm IR}$ and $\beta_{\rm IR}$ parameter values obtained from the fit to observations (red), interpolated (blue) and extrapolated (orange) luminosity functions from other redshifts and transitions. We also show linear regressions using the same color code (the black line shows the fit to all values); all fits have $R^2>0.989$. As it can be appreciated, the values obtained in the interpolation and extrapolations agree well with those from observations.}
     \label{fig:alpha_beta}
 \end{figure}

For those transitions at 150 GHz for which there are available measurements at 90 and 220 GHz (i.e., from $J=2$ up to $J=5$), we choose $\alpha_{\rm IR}$-$\beta_{\rm IR}$ values so that the resulting luminosity function at 150 GHz -- obtained following the same procedure as described in \S~\ref{subsec:COLF} -- lies between the luminosity functions for the same transition at 90 GHz and 220 GHz. This should be an acceptable estimate, since  for $J<4$ the corresponding redshifts at 150 GHz are below the redshifts at which  the cosmic SFR density peaks, and therefore we can expect a monotonic evolution for the CO luminosity. Transitions with $J=4$ and $J=5$ are more subtle, since the redshifts of interest are around the peak cosmic SFR density. In addition, the luminosity functions of these transitions at 90 GHz are peculiar when compared with the rest -- they are the only ones for which the Kamenetzky~\cite{2016ApJ...829...93K} fit values are ruled out by the data and they may require two Schechter functions to fit observations. We refer to this set of choices as \textit{interpolated}.

In turn, there are no available measurements for the $(6\to5)$ and $(7\to6)$ transitions at 90 GHz either, such that we lack any anchor at higher redshifts for the luminosity function. Here, we use trends of the ratios between the luminosity functions at 220 GHz and the fact that the redshift ranges probed at 150 GHz and 90 GHz are above those for which the cosmic SFR density peaks to choose the $\alpha_{\rm IR}$-$\beta_{\rm IR}$ values. Thus, we refer to these values as \textit{extrapolated}. Note that, given how high the redshift this emission comes from is, we expect their contribution to be negligible in any case, as we find in Fig.~\ref{fig:fidclustering}.

We report the whole set of $\alpha_{\rm IR}$-$\beta_{\rm IR}$ values in Table~\ref{tab:alphabeta}, and show a visual representation of them in Fig.~\ref{fig:alpha_beta}. We can see how the values are highly correlated, and that the three kind of sets -- those that we have observations to fit to, interpolated, and extrapolated -- show a very similar correlation. Furthermore, all interpolated and extrapolated cases are fairly consistent with the fits reported in Ref.~\cite{2016ApJ...829...93K} using local measurements.

We therefore acknowledge that our predictions at 150 GHz rely on the interpolated and extrapolated luminosity functions. However, we believe that until there are observations at 150 GHz to calibrate any model to, even an ab initio model that self-consistently predicts all transitions at all redshifts may be inaccurate. We attempt to capture at least part of this uncertainty with the differences between the shallow and steep scenarios.

 \section{The structure of cross shot-noise between $L_{\rm IR}$ and $L_{\rm CO}$}
 \label{appendix:cross-shotnoise}
 If we are considering, simultaneously, the cross shot-noise from IR flux (`1') and CO flux (`2') coming from the same emitter, then expression is given by
 \begin{align}
          C_{\ell}^{\rm \times shot} &= \int dS_1 dS_2 \frac{dN}{dS_1 dS_2} S_1 S_2 \\ 
          &= \int dz dS_1 dS_2 \frac{dN}{dS_1 dS_2 dz} S_1 S_2.
 \end{align}
 Here, the redshift integral in principle runs from $0 \leq z < \infty$, as an IR SED can be arbitrarily redshifted into the bandpass of an experiment. However, since we are also considering line emission through this bandpass the integral bounds are the same as in the CO auto shot-noise case. The IR flux of a source through a CMB experiment is written as 
 \begin{equation}
     S_{\rm IR}(z)  = \frac{L_{\rm IR}}{ 4\pi \chi^2 (z) (1 + z)} \frac{\int d\nu \tau(\nu)\Phi_{\rm IR}(\nu ( 1 + z))}{\int d\nu \tau(\nu)},
 \end{equation}
 where $\tau(\nu)$ is the experimental bandpass and $\Phi_{\rm IR} (\nu) $ the modified black-body SED used to describe IR luminosity. The frequency-dependent contribution of this flux can be re-written as the average of the redshifted SED, $\bar{\Phi}_{\rm IR}(z) \equiv \int d\nu \tau(\nu) \Phi_{\rm IR}(\nu ( 1 + z)) /  \int d\nu \tau(\nu)$. Following arguments analogous to those used to derive the auto shot-noise in \S~\ref{sec:discussion}, we arrive at the form for the cross-shot-noise 
 \begin{align}
     C_\ell^{\times \rm shot} = \frac{c}{4\pi \Delta \nu} \int_{z_{\rm min}}^{z_{\rm max}} \frac{dz}{H(z)} dL_1 dL_2  \frac{d^3 n}{dz dL_1 dL_2 } \frac{ \bar{\Phi}_{\rm IR} (z) L_1 L_2}{\chi^2 (z) (1 + z)^2}.
 \end{align}
The conditional luminosity function, that is, the number density of sources that are simultaneously at redshift $z$ with IR luminosity $L_1$ \emph{and} CO luminosity $L_2$, can be rewritten as a probability distribution function as $d^3n / dz dL_1 dL_2 = \bar{n}(z) P(L_1, L_2 | z) \,$.  
 Adopting Eqn.~\ref{eqn:loglirco} we may directly write $\bar{L}_2 (L_1) = 10^{-\beta_{\rm IR}/\alpha_{\rm IR}} L_1^{1/\alpha_{\rm IR}}$, and we may additionally leverage the fact that we assumed a log-Normal distribution between $L_2$ and $L_1$ to write 
 \begin{align}
     P(L_1, L_2 | z) &= P(L_2 | L_1, z) P(L_1 | z) \\
     &= P(L_1 | z) \frac{1}{L_2 \sqrt{2\pi \sigma^2}} \exp \left ( - \frac{ (\ln L_2 - \ln L_1 / \alpha_{\rm IR} + \ln(10) \beta_{\rm IR} /\alpha_{\rm IR})^2 }{2 \sigma^2} \right ),
 \end{align}
 where $\sigma^2$ is the mean-preserving lognormal scatter, rescaled to the natural logarithm base. A prediction for the number density of CO emitting galaxies $\bar{n}(z)$, and a specific distribution for IR luminosities at a redshift $z$, $P(L_1 | z)$, are the remaining ingredients required to specify an analytic model of the cross shot-noise. \par 
 Assuming the upper integration limit in $L_2$ is infinite\footnote{This is technically not true if the CMB map has a flux cut. In that case the integral over $S_2$ runs until $S_{2,\rm max}$ and this converts to an upper bound on luminosity for every redshift. The resulting integral is the same up to an additional factor of $$\frac{1}{2} \left [ 1 - {\rm erf} \left ( \frac{\sigma^2 + \ln L_{\rm IR}/\alpha - \ln L_{2, \rm max}}{\sqrt{2}\sigma} \right ) \right ], $$ where we have that $L_{2, \rm max}(z) = 4\pi \chi^2 (1+z) S_{2,\rm max} / \bar{\Phi}_{\rm IR}(z)$} we may integrate over $L_2$ and find 
  \begin{align}
  \label{eqn:crosshot}
     C_\ell^{\times \rm shot} = \frac{10^{\sigma^2/2 - \ln(10) \beta_{\rm IR} /\alpha_{\rm IR}}\, c }{4\pi \Delta \nu} \int_{z_{\rm min}}^{z_{\rm max}} \frac{dz\, \bar{\Phi}_{\rm IR} (z)}{H(z) \chi^2 (z) (1 + z)^2} dL_{\rm IR} \frac{d^2 n}{dL_{\rm IR} dz} L_{\rm IR}^{1 + 1/\alpha_{\rm IR}}.
 \end{align}
Given all of these ingredients, we see the cross-shot noise depends on the luminosity function of IR luminosities, and we see that a redshift-weighted moment of the form $\langle L_{\rm IR}^{1 + 1/\alpha} \rangle$ drives the observed cross-shot noise, as discussed in the main text. \par 
 Perhaps interestingly, despite having the highest cross shot-noise we see that $\alpha_{\rm IR}$ for the CO$(4\to 3)$ transition at low frequencies has the highest value -- implying a potentially lower shot noise than if $\alpha_{\rm IR}$ were low. However, since $\alpha_{\rm IR}$ and $\beta_{\rm IR}$ are highly anticorrelated (see Fig.~\ref{fig:alpha_beta}) a large value of $\alpha_{\rm IR}$ begets a highly negative value of $\beta_{\rm IR}$. The amplitude of the shot noise depends on a pre-factor of $10^{-\beta_{\rm IR} / \alpha_{\rm IR}}$, relating $L_{\rm CO}$ to $L_{\rm IR}$. This will compensate for the moment probed not being as large. 
\section{Assessing scales at which CO emission is Poisson dominated}
\label{appendix:poisson}
\begin{figure}
    \centering
    \includegraphics[width=\columnwidth]{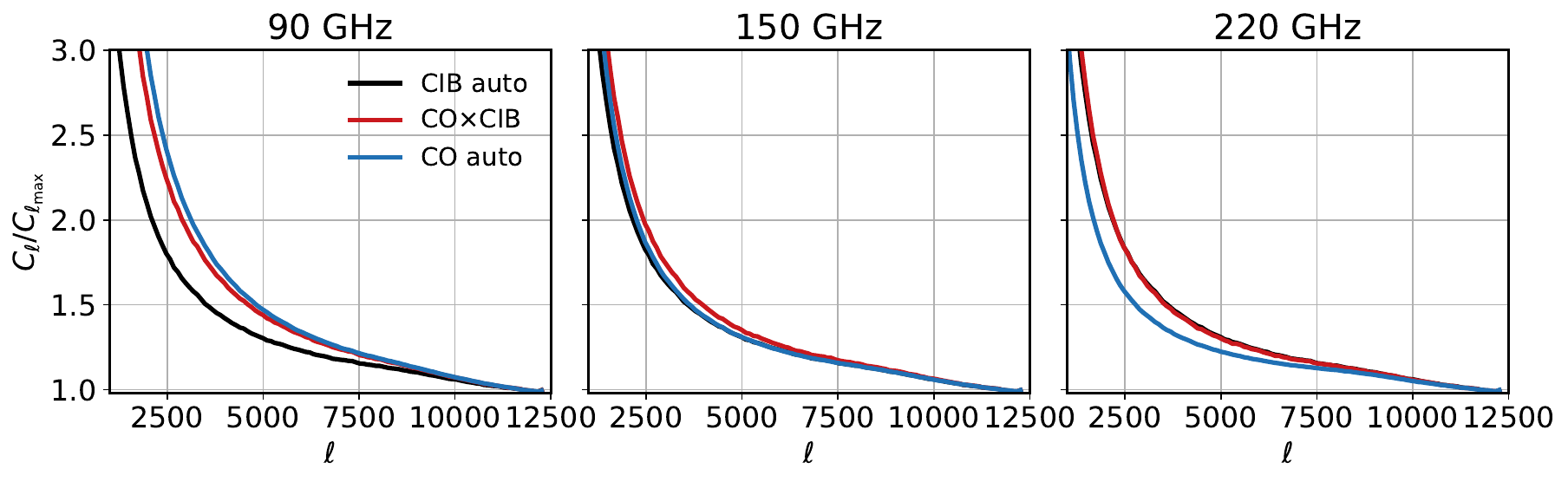}
    \caption{Ratio of $C_\ell$ to the largest angular multipole measured for each frequency band / tracer combination. Under the assumption $C_{\ell_{\rm max}}$ is dominated by the Poisson shot noise component, this ratio measures the relative contributions of clustering to Poisson. We see that all tracers begin to be dominated by the Poisson component (i.e., $C_\ell/C_{\ell_{\rm max}}<2$) at $\ell > 3000$. }
    \label{fig:ellmaxplot}
\end{figure}
The analytic scaling arguments presented in \S~\ref{sec:discussion} relied on the assumption that the measured signal was driven primarily by Poisson shot noise and not by any clustering components. It was discussed that $\ell \sim 3000$ corresponded to a scale at which both CO$\times$CO and the CO$\times$CIB spectrum were dominated by the flat, shot-noise component, assumed to be Poissonian. The purpose of this appendix is to lend credence to the claim regarding the dominant contribution of the shot noise and assess it more carefully. \par 
In lieu of fully modeling a clustered and a Poisson component of halos jointly emitting CO and IR photons, we will adopt an approximate empirical test of the scale at which a Poisson shot noise component begins to dominate the signal in question. Suppose any signal we measured can be broken into an $\ell$-dependent clustering component, monotonically decreasing and a flat shot-noise contribution. Then, we may write it as 
\begin{equation*}
    C_{\ell}^{\rm Total} = C_{\ell}^{\rm clust} + {\rm SN}, 
\end{equation*}
where ${\rm SN}$ is the asymptotic 
shot-noise component whose amplitude scales as Eqn.~\ref{eqn:crosshot}. In the limit of $\ell \to \infty$ the clustering component should vanish, and we should have that $C_{\ell}^{\rm Total} \approx {\rm SN}$. A measurement of the CO$\times$CO or CO$\times$CIB cross-correlation at very high $\ell$, then, should mostly probe the shot-noise component. Dividing the measured $\hat{C}_\ell$ by the highest measured multipole is a rough estimator of 
\begin{equation}
\label{eqn:ellmaxratio}
    \frac{\hat{C}_{\ell}}{\hat{C}_{\ell_{\rm max}}} \approx 1 + \frac{C_{\ell}^{\rm clust}}{\rm SN}.
\end{equation}
Thus, when $\hat{C}_{\ell}/\hat{C}_{\ell_{\rm max}} < 2$, the Poisson component is dominating over the clustering component. \par 
In Fig.~\ref{fig:ellmaxplot} we show the ratios of Eqn.~\ref{eqn:ellmaxratio}, for the CIB auto, CO auto and CO$\times$CIB correlation. The maps were produced at a resolution of $N_{\rm side} = 4096$ which corresponds to $\ell_{\rm max} = 3 N_{\rm side} - 1 = 12287$. We see that the CIB components are nearly always dominated by their 
shot-noise contributions at $\ell > 2000$ for all frequencies. On the other hand, both CO auto and CO$\times$CIB begin to be 
shot-noise dominated at slightly smaller scales of $\ell > 3000$. Therefore, our qualitative analysis assuming only the Poisson component is partially justified, but a full halo model taking a clustering component into account should be constructed. At the best-measured multipole of $\ell = 3000$, for many of these frequencies, both components are comparable. \par 
For 150 and 220GHz, the angular scaling of the relative clustering / shot-noise contributions track those of the \agora CIB quite closely. This implies that the relative clustering and shot-noise components of these models could perhaps be described with an equivalent shape to the combination of the clustering and shot-noise components of the CIB, however this breaks down at 90GHz. In full detail this will also probably not hold if one varies, simultaneously, the amplitudes of the dominant CO lines at every frequency / cross-frequency, but these approximate templates could be valuable in constraining the potential existence of a CO component in current data.

\bibliography{cocibbib.bib}

\providecommand{\href}[2]{#2}\begingroup\raggedright\begin{thebibliography}{10}

\bibitem{2020JCAP...12..047A}
S.~{Aiola}, E.~{Calabrese}, L.~{Maurin}, S.~{Naess}, B.~L. {Schmitt}, M.~H. {Abitbol}, {\em et~al.}, ``{The Atacama Cosmology Telescope: DR4 maps and cosmological parameters},'' \href{http://dx.doi.org/10.1088/1475-7516/2020/12/047}{{\em Journal of Cosmology and Astroparticle Physics} {\bfseries 2020} no.~12, (Dec., 2020) 047}, \href{http://arxiv.org/abs/2007.07288}{{\ttfamily arXiv:2007.07288 [astro-ph.CO]}}.

\bibitem{10.1063/1.3292381}
C.~L. Chang, P.~A.~R. Ade, K.~A. Aird, B.~A. Benson, L.~E. Bleem, J.~E. Carlstrom, {\em et~al.}, ``{SPT‐SZ: a Sunyaev‐ZePdovich survey for galaxy clusters},'' \href{http://dx.doi.org/10.1063/1.3292381}{{\em AIP Conference Proceedings} {\bfseries 1185} no.~1, (12, 2009) 475--477}, \href{http://arxiv.org/abs/https://pubs.aip.org/aip/acp/article-pdf/1185/1/475/12248465/475\_1\_online.pdf}{{\ttfamily https://pubs.aip.org/aip/acp/article-pdf/1185/1/475/12248465/475\_1\_online.pdf}}. \url{https://doi.org/10.1063/1.3292381}.

\bibitem{Austermann_2012}
J.~E. Austermann, K.~A. Aird, J.~A. Beall, D.~Becker, A.~Bender, B.~A. Benson, {\em et~al.}, \href{http://dx.doi.org/10.1117/12.927286}{``SPTpol: an instrument for CMB polarization measurements with the South Pole Telescope,''} in {\em Millimeter, Submillimeter, and Far-Infrared Detectors and Instrumentation for Astronomy VI}, W.~S. Holland, ed.
\newblock SPIE, Sept., 2012.
\newblock \url{http://dx.doi.org/10.1117/12.927286}.

\bibitem{Ade_2019}
P.~Ade, J.~Aguirre, Z.~Ahmed, S.~Aiola, A.~Ali, D.~Alonso, {\em et~al.}, ``The Simons Observatory: science goals and forecasts,'' \href{http://dx.doi.org/10.1088/1475-7516/2019/02/056}{{\em Journal of Cosmology and Astroparticle Physics} {\bfseries 2019} no.~02, (Feb., 2019) 056–056}. \url{http://dx.doi.org/10.1088/1475-7516/2019/02/056}.

\bibitem{CCAT_Prime_Collaboration_2022}
C.-P. Collaboration, M.~Aravena, J.~E. Austermann, K.~Basu, N.~Battaglia, B.~Beringue, {\em et~al.}, ``CCAT-prime Collaboration: Science Goals and Forecasts with Prime-Cam on the Fred Young Submillimeter Telescope,'' \href{http://dx.doi.org/10.3847/1538-4365/ac9838}{{\em The Astrophysical Journal Supplement Series} {\bfseries 264} no.~1, (Dec., 2022) 7}. \url{http://dx.doi.org/10.3847/1538-4365/ac9838}.

\bibitem{Reichardt_2012}
C.~L. Reichardt, L.~Shaw, O.~Zahn, K.~A. Aird, B.~A. Benson, L.~E. Bleem, {\em et~al.}, ``A MEASUREMENT OF SECONDARY COSMIC MICROWAVE BACKGROUND ANISOTROPIES WITH TWO YEARS OF SOUTH POLE TELESCOPE OBSERVATIONS,'' \href{http://dx.doi.org/10.1088/0004-637x/755/1/70}{{\em The Astrophysical Journal} {\bfseries 755} no.~1, (July, 2012) 70}. \url{http://dx.doi.org/10.1088/0004-637X/755/1/70}.

\bibitem{Dunkley_2013}
J.~Dunkley, E.~Calabrese, J.~Sievers, G.~Addison, N.~Battaglia, E.~Battistelli, {\em et~al.}, ``The Atacama Cosmology Telescope: likelihood for small-scale CMB data,'' \href{http://dx.doi.org/10.1088/1475-7516/2013/07/025}{{\em Journal of Cosmology and Astroparticle Physics} {\bfseries 2013} no.~07, (July, 2013) 025–025}. \url{http://dx.doi.org/10.1088/1475-7516/2013/07/025}.

\bibitem{PlanckXIII}
P.~A.~R. Ade, N.~Aghanim, M.~Arnaud, M.~Ashdown, J.~Aumont, C.~Baccigalupi, {\em et~al.}, ``Planck2015 results: XIII. Cosmological parameters,'' \href{http://dx.doi.org/10.1051/0004-6361/201525830}{{\em Astronomy \& Astrophysics} {\bfseries 594} (Sept., 2016) A13}. \url{http://dx.doi.org/10.1051/0004-6361/201525830}.

\bibitem{Reichardt_2021}
C.~L. Reichardt, S.~Patil, P.~A.~R. Ade, A.~J. Anderson, J.~E. Austermann, J.~S. Avva, {\em et~al.}, ``An Improved Measurement of the Secondary Cosmic Microwave Background Anisotropies from the SPT-SZ + SPTpol Surveys,'' \href{http://dx.doi.org/10.3847/1538-4357/abd407}{{\em The Astrophysical Journal} {\bfseries 908} no.~2, (Feb., 2021) 199}. \url{http://dx.doi.org/10.3847/1538-4357/abd407}.

\bibitem{Universemachine}
P.~{Behroozi}, R.~H. {Wechsler}, A.~P. {Hearin}, and C.~{Conroy}, ``{UNIVERSEMACHINE: The correlation between galaxy growth and dark matter halo assembly from z = 0-10},'' \href{http://dx.doi.org/10.1093/mnras/stz1182}{{\em Monthly Notices of the Royal Astronomical Society} {\bfseries 488} no.~3, (Sept., 2019) 3143--3194}, \href{http://arxiv.org/abs/1806.07893}{{\ttfamily arXiv:1806.07893 [astro-ph.GA]}}.

\bibitem{Righi_2008}
M.~Righi, C.~Hernández-Monteagudo, and R.~A. Sunyaev, ``Carbon monoxide line emission as a CMB foreground: tomography of the star-forming universe with different spectral resolutions,'' \href{http://dx.doi.org/10.1051/0004-6361:200810199}{{\em Astronomy \&; Astrophysics} {\bfseries 489} no.~2, (July, 2008) 489–504}. \url{http://dx.doi.org/10.1051/0004-6361:200810199}.

\bibitem{Bernal:2022jap}
J.~L. Bernal and E.~D. Kovetz, ``{Line-intensity mapping: theory review with a focus on star-formation lines},'' \href{http://dx.doi.org/10.1007/s00159-022-00143-0}{{\em Astron. Astrophys. Rev.} {\bfseries 30} no.~1, (2022) 5}, \href{http://arxiv.org/abs/2206.15377}{{\ttfamily arXiv:2206.15377 [astro-ph.CO]}}.

\bibitem{Maniyar_2023}
A.~S. Maniyar, A.~Gkogkou, W.~R. Coulton, Z.~Li, G.~Lagache, and A.~R. Pullen, ``Extragalactic CO emission lines in the CMB experiments: A forgotten signal and a foreground,'' \href{http://dx.doi.org/10.1103/physrevd.107.123504}{{\em Physical Review D} {\bfseries 107} no.~12, (June, 2023) }. \url{http://dx.doi.org/10.1103/PhysRevD.107.123504}.

\bibitem{omori2022agora}
Y.~Omori, ``Agora: Multi-Component Simulation for Cross-Survey Science,'' 2022.

\bibitem{skyline}
G.~Sato-Polito, N.~Kokron, and J.~L. Bernal, ``A multitracer empirically driven approach to line-intensity mapping light cones,'' \href{http://dx.doi.org/10.1093/mnras/stad2498}{{\em Monthly Notices of the Royal Astronomical Society} {\bfseries 526} no.~4, (Aug., 2023) 5883–5899}. \url{http://dx.doi.org/10.1093/mnras/stad2498}.

\bibitem{bolliet2023classsz}
B.~Bolliet, A.~Kusiak, F.~McCarthy, A.~Sabyr, K.~Surrao, J.~C. Hill, {\em et~al.}, ``class\_sz I: Overview,'' 2023.

\bibitem{Bernal_2019}
J.~L. Bernal, P.~C. Breysse, H.~Gil-Marín, and E.~D. Kovetz, ``User’s guide to extracting cosmological information from line-intensity maps,'' \href{http://dx.doi.org/10.1103/physrevd.100.123522}{{\em Physical Review D} {\bfseries 100} no.~12, (Dec., 2019) }. \url{http://dx.doi.org/10.1103/PhysRevD.100.123522}.

\bibitem{2019ApJ...887..142S}
G.~{Sun}, B.~S. {Hensley}, T.-C. {Chang}, O.~{Dor{\'e}}, and P.~{Serra}, ``{A Self-consistent Framework for Multiline Modeling in Line Intensity Mapping Experiments},'' \href{http://dx.doi.org/10.3847/1538-4357/ab55df}{{\em \apj} {\bfseries 887} no.~2, (Dec., 2019) 142}, \href{http://arxiv.org/abs/1907.02999}{{\ttfamily arXiv:1907.02999 [astro-ph.GA]}}.

\bibitem{Schaan_2021}
E.~Schaan and M.~White, ``Multi-tracer intensity mapping: cross-correlations, line noise \&; decorrelation,'' \href{http://dx.doi.org/10.1088/1475-7516/2021/05/068}{{\em Journal of Cosmology and Astroparticle Physics} {\bfseries 2021} no.~05, (May, 2021) 068}. \url{http://dx.doi.org/10.1088/1475-7516/2021/05/068}.

\bibitem{Stein_2020}
G.~Stein, M.~A. Alvarez, J.~R. Bond, A.~v. Engelen, and N.~Battaglia, ``The Websky extragalactic CMB simulations,'' \href{http://dx.doi.org/10.1088/1475-7516/2020/10/012}{{\em Journal of Cosmology and Astroparticle Physics} {\bfseries 2020} no.~10, (Oct., 2020) 012–012}. \url{http://dx.doi.org/10.1088/1475-7516/2020/10/012}.

\bibitem{B_thermin_2017}
M.~Béthermin, H.-Y. Wu, G.~Lagache, I.~Davidzon, N.~Ponthieu, M.~Cousin, {\em et~al.}, ``The impact of clustering and angular resolution on far-infrared and millimeter continuum observations,'' \href{http://dx.doi.org/10.1051/0004-6361/201730866}{{\em Astronomy \&; Astrophysics} {\bfseries 607} (Nov., 2017) A89}. \url{http://dx.doi.org/10.1051/0004-6361/201730866}.

\bibitem{B_thermin_2022}
M.~Béthermin, A.~Gkogkou, M.~Van~Cuyck, G.~Lagache, A.~Beelen, M.~Aravena, {\em et~al.}, ``CONCERTO: High-fidelity simulation of millimeter line emissions of galaxies and [CII] intensity mapping,'' \href{http://dx.doi.org/10.1051/0004-6361/202243888}{{\em Astronomy \&; Astrophysics} {\bfseries 667} (Nov., 2022) A156}. \url{http://dx.doi.org/10.1051/0004-6361/202243888}.

\bibitem{MDPL2}
A.~Klypin, G.~Yepes, S.~Gottlober, F.~Prada, and S.~Hess, ``{MultiDark simulations: the story of dark matter halo concentrations and density profiles},'' \href{http://dx.doi.org/10.1093/mnras/stw248}{{\em Mon. Not. Roy. Astron. Soc.} {\bfseries 457} no.~4, (2016) 4340--4359}, \href{http://arxiv.org/abs/1411.4001}{{\ttfamily arXiv:1411.4001 [astro-ph.CO]}}.

\bibitem{Rockstar}
P.~S. {Behroozi}, R.~H. {Wechsler}, and H.-Y. {Wu}, ``{The ROCKSTAR Phase-space Temporal Halo Finder and the Velocity Offsets of Cluster Cores},'' \href{http://dx.doi.org/10.1088/0004-637X/762/2/109}{{\em \apj} {\bfseries 762} no.~2, (Jan., 2013) 109}, \href{http://arxiv.org/abs/1110.4372}{{\ttfamily arXiv:1110.4372 [astro-ph.CO]}}.

\bibitem{Schaan_2018}
E.~Schaan, S.~Ferraro, and D.~N. Spergel, ``Weak lensing of intensity mapping: The cosmic infrared background,'' \href{http://dx.doi.org/10.1103/physrevd.97.123539}{{\em Physical Review D} {\bfseries 97} no.~12, (June, 2018) }. \url{http://dx.doi.org/10.1103/PhysRevD.97.123539}.

\bibitem{Carilli:2013qm}
C.~Carilli and F.~Walter, ``{Cool Gas in High Redshift Galaxies},'' \href{http://dx.doi.org/10.1146/annurev-astro-082812-140953}{{\em Ann. Rev. Astron. Astrophys.} {\bfseries 51} (2013) 105--161}, \href{http://arxiv.org/abs/1301.0371}{{\ttfamily arXiv:1301.0371 [astro-ph.CO]}}.

\bibitem{2016ApJ...829...93K}
J.~{Kamenetzky}, N.~{Rangwala}, J.~{Glenn}, P.~R. {Maloney}, and A.~{Conley}, ``{L$_{CO}$/L$_{FIR}$ Relations with CO Rotational Ladders of Galaxies Across the Herschel SPIRE Archive},'' \href{http://dx.doi.org/10.3847/0004-637X/829/2/93}{{\em \apj} {\bfseries 829} no.~2, (Oct., 2016) 93}, \href{http://arxiv.org/abs/1508.05102}{{\ttfamily arXiv:1508.05102 [astro-ph.GA]}}.

\bibitem{2012ARA&A..50..531K}
R.~C. {Kennicutt} and N.~J. {Evans}, ``{Star Formation in the Milky Way and Nearby Galaxies},'' \href{http://dx.doi.org/10.1146/annurev-astro-081811-125610}{{\em Annual Reviews of Astronomy and Astrophysics} {\bfseries 50} (Sept., 2012) 531--608}, \href{http://arxiv.org/abs/1204.3552}{{\ttfamily arXiv:1204.3552 [astro-ph.GA]}}.

\bibitem{MadauDickinson}
P.~Madau and M.~Dickinson, ``{Cosmic Star Formation History},'' \href{http://dx.doi.org/10.1146/annurev-astro-081811-125615}{{\em Ann. Rev. Astron. Astrophys.} {\bfseries 52} (2014) 415--486}, \href{http://arxiv.org/abs/1403.0007}{{\ttfamily arXiv:1403.0007 [astro-ph.CO]}}.

\bibitem{Bouwens2020}
R.~{Bouwens}, J.~{Gonz{\'a}lez-L{\'o}pez}, M.~{Aravena}, R.~{Decarli}, M.~{Novak}, M.~{Stefanon}, {\em et~al.}, ``{The ALMA Spectroscopic Survey Large Program: The Infrared Excess of z = 1.5-10 UV-selected Galaxies and the Implied High-redshift Star Formation History},'' \href{http://dx.doi.org/10.3847/1538-4357/abb830}{{\em \apj} {\bfseries 902} no.~2, (Oct., 2020) 112}, \href{http://arxiv.org/abs/2009.10727}{{\ttfamily arXiv:2009.10727 [astro-ph.GA]}}.

\bibitem{hogg2000distance}
D.~W. Hogg, ``Distance measures in cosmology,'' 2000.

\bibitem{2020ApJ...902..110D}
R.~{Decarli}, M.~{Aravena}, L.~{Boogaard}, C.~{Carilli}, J.~{Gonz{\'a}lez-L{\'o}pez}, F.~{Walter}, {\em et~al.}, ``{The ALMA Spectroscopic Survey in the Hubble Ultra Deep Field: Multiband Constraints on Line-luminosity Functions and the Cosmic Density of Molecular Gas},'' \href{http://dx.doi.org/10.3847/1538-4357/abaa3b}{{\em \apj} {\bfseries 902} no.~2, (Oct., 2020) 110}, \href{http://arxiv.org/abs/2009.10744}{{\ttfamily arXiv:2009.10744 [astro-ph.GA]}}.

\bibitem{Boogaard:2023whs}
L.~A. Boogaard {\em et~al.}, ``{A NOEMA Molecular Line Scan of the Hubble Deep Field North: Improved Constraints on the CO Luminosity Functions and Cosmic Density of Molecular Gas},'' \href{http://dx.doi.org/10.3847/1538-4357/acb4f0}{{\em Astrophys. J.} {\bfseries 945} no.~2, (2023) 111}, \href{http://arxiv.org/abs/2301.05705}{{\ttfamily arXiv:2301.05705 [astro-ph.GA]}}.

\bibitem{2020ApJ...895...81R}
D.~A. {Riechers}, J.~A. {Hodge}, R.~{Pavesi}, E.~{Daddi}, R.~{Decarli}, R.~J. {Ivison}, {\em et~al.}, ``{COLDz: A High Space Density of Massive Dusty Starburst Galaxies {\ensuremath{\sim}}1 Billion Years after the Big Bang},'' \href{http://dx.doi.org/10.3847/1538-4357/ab8c48}{{\em \apj} {\bfseries 895} no.~2, (June, 2020) 81}, \href{http://arxiv.org/abs/2004.10204}{{\ttfamily arXiv:2004.10204 [astro-ph.GA]}}.

\bibitem{Li_2016}
T.~Y. Li, R.~H. Wechsler, K.~Devaraj, and S.~E. Church, ``CONNECTING CO INTENSITY MAPPING TO MOLECULAR GAS AND STAR FORMATION IN THE EPOCH OF GALAXY ASSEMBLY,'' \href{http://dx.doi.org/10.3847/0004-637x/817/2/169}{{\em The Astrophysical Journal} {\bfseries 817} no.~2, (Jan., 2016) 169}. \url{http://dx.doi.org/10.3847/0004-637X/817/2/169}.

\bibitem{Sun_2021}
G.~Sun, T.-C. Chang, B.~D. Uzgil, J.~J. Bock, C.~M. Bradford, V.~Butler, {\em et~al.}, ``Probing Cosmic Reionization and Molecular Gas Growth with TIME,'' \href{http://dx.doi.org/10.3847/1538-4357/abfe62}{{\em The Astrophysical Journal} {\bfseries 915} no.~1, (July, 2021) 33}. \url{http://dx.doi.org/10.3847/1538-4357/abfe62}.

\bibitem{comap}
K.~A. {Cleary}, J.~{Borowska}, P.~C. {Breysse}, M.~{Catha}, D.~T. {Chung}, S.~E. {Church}, {\em et~al.}, ``{COMAP Early Science. I. Overview},'' \href{http://dx.doi.org/10.3847/1538-4357/ac63cc}{{\em \apj} {\bfseries 933} no.~2, (July, 2022) 182}, \href{http://arxiv.org/abs/2111.05927}{{\ttfamily arXiv:2111.05927 [astro-ph.CO]}}.

\bibitem{Planck_HFI}
P.~A.~R. Ade, N.~Aghanim, C.~Armitage-Caplan, M.~Arnaud, M.~Ashdown, F.~Atrio-Barandela, {\em et~al.}, ``Planck2013 results. IX. HFI spectral response,'' \href{http://dx.doi.org/10.1051/0004-6361/201321531}{{\em Astronomy \&; Astrophysics} {\bfseries 571} (Oct., 2014) A9}. \url{http://dx.doi.org/10.1051/0004-6361/201321531}.

\bibitem{Gorski_2005}
K.~M. Gorski, E.~Hivon, A.~J. Banday, B.~D. Wandelt, F.~K. Hansen, M.~Reinecke, and M.~Bartelmann, ``HEALPix: A Framework for High‐Resolution Discretization and Fast Analysis of Data Distributed on the Sphere,'' \href{http://dx.doi.org/10.1086/427976}{{\em The Astrophysical Journal} {\bfseries 622} no.~2, (Apr., 2005) 759–771}. \url{http://dx.doi.org/10.1086/427976}.

\bibitem{Moradinezhad_Dizgah_2022b}
A.~Moradinezhad~Dizgah, F.~Nikakhtar, G.~K. Keating, and E.~Castorina, ``Precision tests of CO and [CII] power spectra models against simulated intensity maps,'' \href{http://dx.doi.org/10.1088/1475-7516/2022/02/026}{{\em Journal of Cosmology and Astroparticle Physics} {\bfseries 2022} no.~02, (Feb., 2022) 026}. \url{http://dx.doi.org/10.1088/1475-7516/2022/02/026}.

\bibitem{Obuljen:2022cjo}
A.~Obuljen, M.~Simonovi\'c, A.~Schneider, and R.~Feldmann, ``{Modeling HI at the field level},'' \href{http://dx.doi.org/10.1103/PhysRevD.108.083528}{{\em Phys. Rev. D} {\bfseries 108} no.~8, (2023) 083528}, \href{http://arxiv.org/abs/2207.12398}{{\ttfamily arXiv:2207.12398 [astro-ph.CO]}}.

\bibitem{NIST:DLMF}
``{\it NIST Digital Library of Mathematical Functions}.'' \url{https://dlmf.nist.gov/}, release 1.2.0 of 2024-03-15.
\newblock \url{https://dlmf.nist.gov/}. F.~W.~J. Olver, A.~B. {Olde Daalhuis}, D.~W. Lozier, B.~I. Schneider, R.~F. Boisvert, C.~W. Clark, B.~R. Miller, B.~V. Saunders, H.~S. Cohl, and M.~A. McClain, eds.

\bibitem{Gkogkou:2022bzo}
A.~Gkogkou {\em et~al.}, ``{CONCERTO: Simulating the CO, [CII], and [CI] line emission of galaxies in a 117 deg2 field and the impact of field-to-field variance},'' \href{http://dx.doi.org/10.1051/0004-6361/202245151}{{\em Astron. Astrophys.} {\bfseries 670} (2023) A16}, \href{http://arxiv.org/abs/2212.02235}{{\ttfamily arXiv:2212.02235 [astro-ph.CO]}}.

\bibitem{Kusiak:2021hai}
A.~Kusiak, B.~Bolliet, S.~Ferraro, J.~C. Hill, and A.~Krolewski, ``{Constraining the baryon abundance with the kinematic Sunyaev-Zel\textquoteright{}dovich effect: Projected-field detection using Planck, WMAP, and unWISE},'' \href{http://dx.doi.org/10.1103/PhysRevD.104.043518}{{\em Phys. Rev. D} {\bfseries 104} no.~4, (2021) 043518}, \href{http://arxiv.org/abs/2102.01068}{{\ttfamily arXiv:2102.01068 [astro-ph.CO]}}.

\bibitem{Yang:2019eoj}
S.~Yang, A.~R. Pullen, and E.~R. Switzer, ``{Evidence for C II diffuse line emission at redshift $z\sim2.6$},'' \href{http://dx.doi.org/10.1093/mnrasl/slz126}{{\em Mon. Not. Roy. Astron. Soc.} {\bfseries 489} no.~1, (2019) L53--L57}, \href{http://arxiv.org/abs/1904.01180}{{\ttfamily arXiv:1904.01180 [astro-ph.CO]}}.

\bibitem{2020NumPy-Array}
C.~R. Harris, K.~J. Millman, S.~J. van~der Walt, R.~Gommers, P.~Virtanen, D.~Cournapeau, {\em et~al.}, ``Array programming with {NumPy},'' \href{http://dx.doi.org/10.1038/s41586-020-2649-2}{{\em Nature} {\bfseries 585} (2020) 357–362}.

\bibitem{2020SciPy-NMeth}
P.~Virtanen, R.~Gommers, T.~E. Oliphant, M.~Haberland, T.~Reddy, D.~Cournapeau, {\em et~al.}, ``{{SciPy} 1.0: Fundamental Algorithms for Scientific Computing in Python},'' \href{http://dx.doi.org/10.1038/s41592-019-0686-2}{{\em Nature Methods} {\bfseries 17} (2020) 261--272}.

\bibitem{4160265}
J.~D. {Hunter}, ``Matplotlib: A 2D Graphics Environment,'' \href{http://dx.doi.org/10.1109/MCSE.2007.55}{{\em Computing in Science Engineering} {\bfseries 9} no.~3, (2007) 90--95}.

\end{thebibliography}\endgroup
\bibliographystyle{utcaps}

\end{document}